\documentclass{aa}
\pdfoutput=1

\usepackage[varg]{txfonts}

\usepackage{color}           
\usepackage[dvipsnames]{xcolor}   

\usepackage[inline]{enumitem} 

\usepackage{graphicx}  
\usepackage{subfig}

\newcommand{\kms}{km~s$^{-1}\,$}

\newcommand{\degm}{^{\circ}\,}

\begin{document}

   \title{Emergence of non-twisted magnetic fields in the Sun: Jets and atmospheric response}
   \titlerunning{Emergence of non-twisted magnetic fields in the Sun}

   \author{P. Syntelis
          \inst{1,2}
          \and
          V. Archontis\inst{2,3}
          \and
          C. Gontikakis\inst{1}
          \and
          K. Tsinganos
          \inst{2}
          }

   \institute{Reasearch Center for Astronomy and Applied Mathematics, Academy of Athens, 4 Soranou Efessiou Str. Athens 11527, Greece,
              \email{psyntelis@phys.uoa.gr}
         \and
            Section of Astrophysics, Astronomy and Mechanics, Department of Physics, University of Athens, Panepistimiopolis, Zografos 15784, Athens, Greece
		 \and
		 	School of Mathematics and Statistics, St. Andrews University, St. Andrews, KY16 9SS, UK
             }


 
	\abstract
	{}
	{We study the emergence of a non-twisted flux tube from the solar
	interior into the solar atmosphere. We investigate whether the length of the buoyant part of the 
flux tube (i.e. $\lambda$) affects the emergence of the field and the dynamics of the evolving magnetic flux system.}
	{We perform three-dimensional (3D), time-dependent, resistive, compressible
	MHD simulations using the Lare3D code.}
	{We find that there are considerable differences in the dynamics of the
	emergence of a magnetic flux tube when $\lambda$ is varied.
	In the solar interior, for larger values of $\lambda$, the
	rising magnetic field emerges faster and expands more due to its lower
	magnetic tension.
	As a result, its field strength decreases and its emergence above the
	photosphere occurs later than in the smaller $\lambda$ case.
	However, in both cases, the emerging field at the photosphere becomes
	unstable in two places, forming two magnetic bipoles that interact
	dynamically during the evolution of the system.
	Most of the dynamic phenomena occur at the current layer, which is formed
	at the interface between the interacting bipoles. We find the formation
	and ejection of plasmoids, the onset of successive jets from the interface, and the 
	impulsive heating of the plasma in the solar atmosphere.
	We discuss the triggering mechanism of the jets and the atmospheric
	response to the emergence of magnetic flux in the two cases.}
	{}

   \keywords{Sun: activity -- Sun:interior -- 
                Sun: Magnetic fields --Magnetohydrodynamics (MHD) --methods: numerical
               }

   \maketitle

\section{Introduction}
\begin{figure*}
	\centering
	\includegraphics[width=0.98\textwidth]{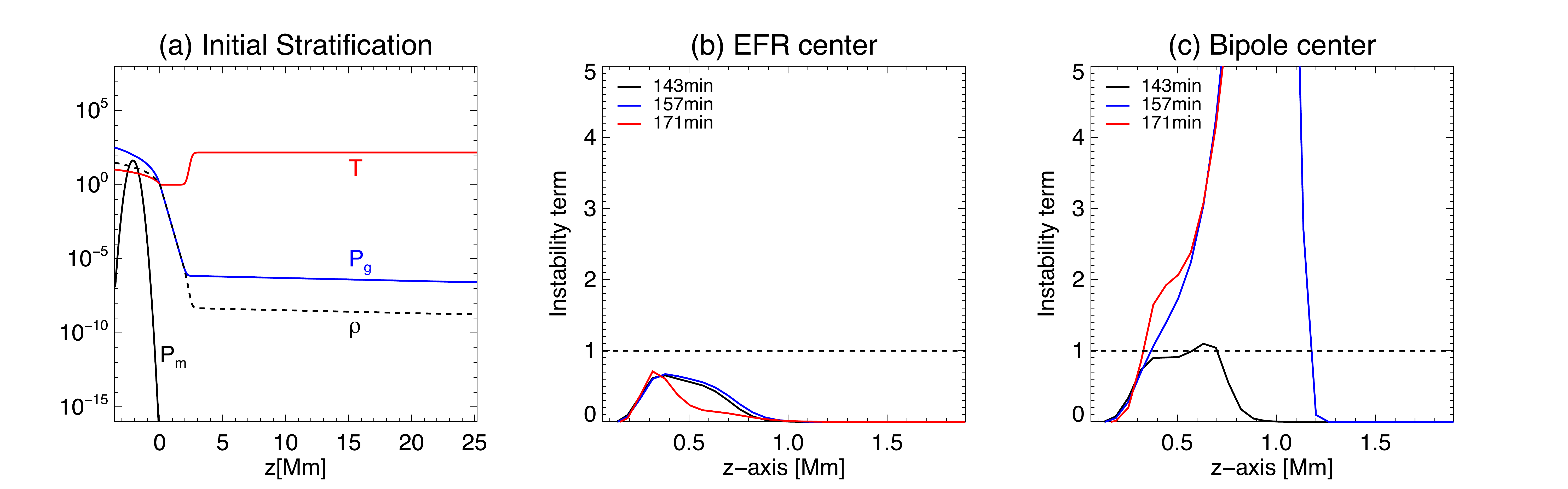}
	\caption{\textbf{(a):} Initial stratification of the atmosphere (temperature (T), density ($\rho$), magnetic pressure ($P_m$), and gas pressure ($P_g$)). \textbf{(b):} Instability term at the center of the EFR and \textbf{(c):} at the center of the bipole at $y=4.8$~Mm.
	}
	\label{fig:stratification_instability}
\end{figure*}

Active region formation and dynamic phenomena on the Sun have been associated with magnetic flux emergence from both a theoretical and observational perspective \citep[e.g.][]{Parker_1955,Tsinganos_1980,Priest_1982,Zwaan_1985,Shibata_1999,Archontis_etal2004,Schrijver_2009}.
In numerical flux emergence models, a typical initial condition for the subphotospheric magnetic field is a twisted magnetic flux tube, which undergoes a buoyant rise towards the solar surface.

The buoyant rise is initiated by removing plasma density along the flux tube axis in a Gaussian manner ($\Delta\rho\propto \exp{(-y^2/\lambda^2)}$). 
The length of the buoyant part of the flux tube (commonly denoted by $\lambda$) is an important parameter that affects the emergence of the flux tube and the following dynamics.
For instance, \citet{Fan_2001} performed an experiment of a flux tube with $\lambda=20$ (i.e. $\approx 3$~Mm), which emerged into the solar atmosphere. \citet{Manchester_etal2004}, used the same initial conditions as \citet{Fan_2001}, but with half the length of the buoyant section of the flux tube ($\lambda=10$). They found that the decrease in $\lambda$ results in more efficient plasma draining. 
In addition, they found the formation of a new flux rope at the low atmosphere due to the shearing and reconnection of the fieldlines across the 
polarity inversion line (PIL).
\citet{Archontis_etal2010} studied the effect of different field strengths, twist, and $\lambda$ to the formation of magnetic tails. For a highly twisted flux tube with high field strength, located at the upper convection zone, they changed the buoyant part of the flux tube from $\lambda$=10 to $\lambda$=20. This resulted in less magnetic tension on field lines of the higher $\lambda$ during the initial emergence, leading to relatively faster emergence. 

Another parameter that plays a major role in the emergence and the dynamics of flux tubes is the twist of the magnetic field. In many twisted flux tube models \citep[e.g.][]{Fan_2001, Magara_etal2003, Archontis_etal2005}, the field lines wind about the axis of the tube in a uniform manner. 
A common setup for such axisymmetric flux tubes is to assume that the flux tube has a component of its magnetic field  
along its axis ($B_{axis}$) and a poloidal component that is proportional to $B_{axis}$ (i.e. $B_{\phi}\propto \alpha B_{axis}$). 
This parameter $\alpha$ measures the twist of the magnetic field per unit of length. It also defines the pitch 
angle between the magnetic field vector and the axis of the flux tube.
Highly twisted flux tubes with pitch angle greater than $45\degm$ would become kink unstable \citep{Linton_etal1996,Fan_etal1999}.

Concerning the ability of low-twist magnetic fields to emerge above the photosphere, simulations have indicated that the flux tubes can become fragmented into two counter-rotating vortices during their emergence \citep{Schuessler_1979}. This fragmentation can be suppressed by the addition of a sufficient amount of twist \citep[e.g.][]{Moreno-Insertis_etal1996}. On the other hand, by performing simulations with different perturbations of specific entropy in the flux tube, \citet{Abbett_etal2000} found that even when the twist of the initial tube is not sufficient, the tube can remain coherent provided it has sufficiently large curvature at its rising apex. \citet{Abbett_etal2001} found that the Coriolis force also suppresses the degree of fragmentation of the apex of the emerging flux tube. 
Also, if the magnetic field is much stronger than the equipartition value of the magnetic field in the convection zone, the buoyancy force dominates over the convective downdrafts, and the flux tube can emerge unimpeded \citep{Fan_etal2003}.
Failed emergence of flux tubes with low twist have been reported in several studies. \citet{Murray_etal2006} performed experiments of flux tubes with $\lambda=20$, for different values of twist ($\alpha=0.1-0.4$) and found that the highly twisted flux tube ($\alpha=0.4$) emerged into the atmosphere forming a bipolar region on the photosphere. Flux tubes with intermediate twist ($\alpha=0.2$) emerged forming two bipolar regions (like the ``sea serpent'' configuration) on the photosphere. Weakly twisted flux tubes ($\alpha=0.1$) failed to emerge above the photosphere. Instead, they continued to expand horizontally inside the solar interior. \citet{Toriumi_etal2011} also studied the emergence of flux tubes with $\lambda=20$ \citep[and similar initial conditions to][]{Murray_etal2006} for a wider range of twists ($\alpha=0.05-0.5$) and found similar results to \citet{Murray_etal2006}. Those results indicate that weakly twisted flux tubes struggle to emerge above the photosphere.

The effect of $\lambda$ (which affects the curvature of the emerging part of flux tube) on the emergence of weakly twisted flux tubes was studied by \citet{Archontis_etal2013}. They used similar initial conditions to \citet{Murray_etal2006} and \citet{Toriumi_etal2011} for a flux tube with $\alpha=0.1$, but assumed half the length of the buoyant part of the flux tube ($\lambda=10$).
In this case, the weakly twisted flux tube did not fail to emerge above the photophere. Instead, the flux tube formed two bipolar regions on the photosphere \citep[similar to the $\alpha=0.2$ case of][]{Murray_etal2006}. Above the two bipoles, the expanding magnetic field formed two magnetic lobes. The interaction of the lobes triggered several reconnection jets and created an envelope field that confined them. The 
shearing motion and reconnection of field lines along the PIL led to the formation of a highly twisted flux rope, which remained at 
the low atmosphere during the running time of the simulation.

In this paper, we advance the work by \citet{Archontis_etal2013} on the emergence of weakly twisted fields by adding two key elements: 
Firstly, we use a non-twisted flux tube and secondly, we study the effect of $\lambda$ on the overall dynamics of the emerging flux system.
The use of a non-twisted flux tube ($\alpha=0$) is chosen because non-twisted fields are more susceptible to distortion and/or fragmentation and, consequently, 
to an overall failed emergence. Furthermore, we want to examine whether the emergence of a non-twisted magnetic field 
can create twisted magnetic structures in the solar atmosphere, similar to the post-emergence 
flux ropes formed at the weakly twisted case by \citet{Archontis_etal2013}.

This paper is organized in the following manner.
In Section \ref{sec:initial_conditions} we describe the initial conditions of the experiments. In Section \ref{sec:lambda_10}, we study the case of $\lambda=10$ in terms of its initial emergence, the dynamics of the interaction between the formed magnetic bipoles, the coronal response, and the energy and flux transfer into the atmosphere. In Section \ref{sec:lambda_5}, we study the $\lambda=5$ case with respect to the same aspects and compare the results with the $\lambda=10$ case. 
In Section \ref{sec:conclusions}, we summarize the results.


\section{Initial conditions and numerical code}

\begin{figure*}
	\centering
	\includegraphics[width=0.85\textwidth]{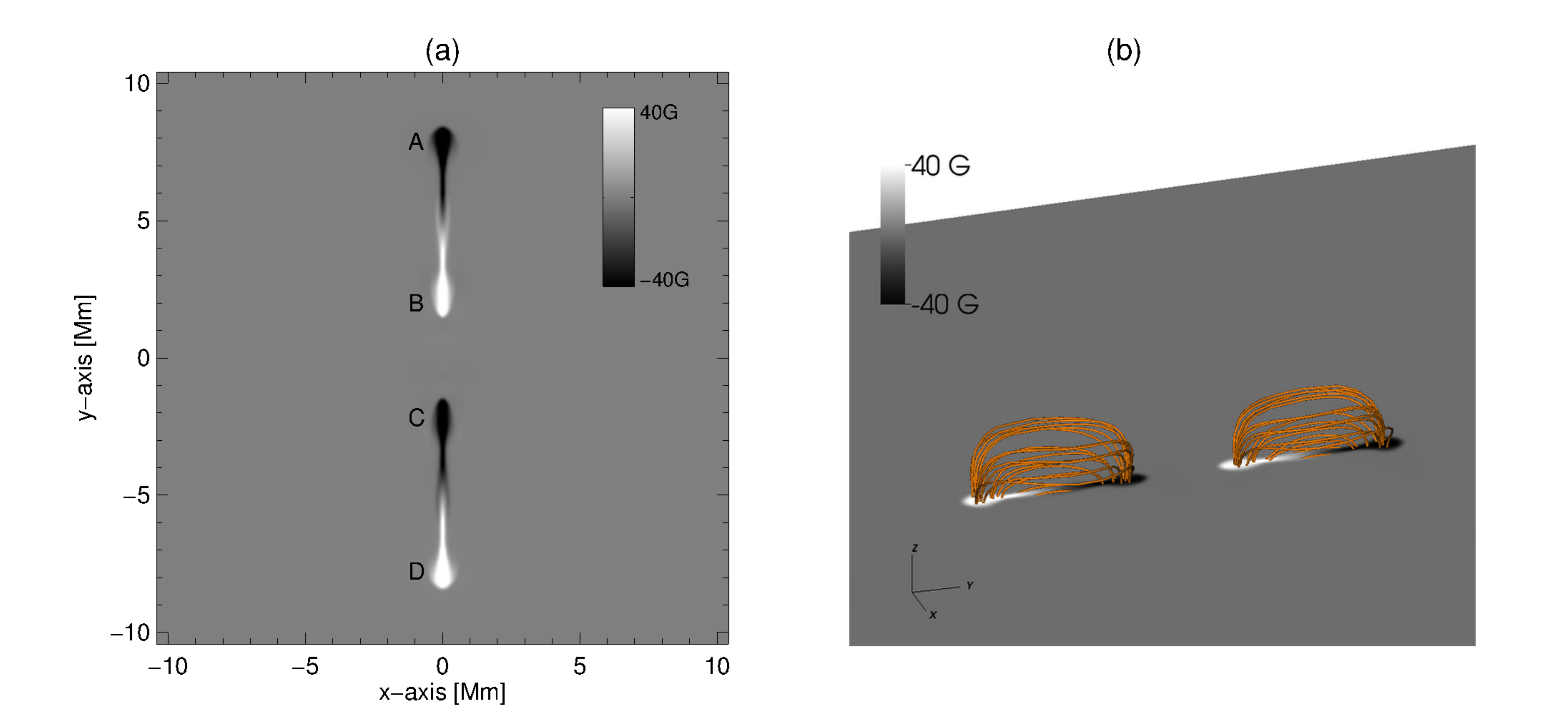}
	\caption{\textbf{(a):} Magnetogram at the mid-photosphere ($z=0.8$~Mm) at $t=169$~min. For simplicity, we call A,D the outer polarities and B,C the inner polarities.
	\textbf{(b):} Field lines traced from the bipoles at $t=169$~min. 
	\label{fig:magneto_tail}
	}
\end{figure*}


\label{sec:initial_conditions}
In order to study the emergence and evolution of the non-twisted flux tube, we solve the 3D time-dependent, resistive, compressible MHD equations in Cartesian geometry. We use the Lare3D code of \citet{Arber_etal2001} to solve numerically the equations. The equations in dimensionless form are
\begin{align}
&\frac{\partial \rho}{\partial t}+ \nabla \cdot (\rho \mathbf{v})  =0 ,\\
&\frac{\partial (\rho \mathbf{v})}{\partial t}  = - \nabla \cdot (\rho \mathbf{v v})  + (\nabla \times \mathbf{B}) \times \mathbf{B} - \nabla P + \rho \mathbf{g} + \nabla \cdot \mathbf{S} , \\
&\frac{ \partial ( \rho \epsilon )}{\partial t} = - \nabla \cdot (\rho \epsilon \mathbf{v}) -P \nabla \cdot \mathbf{v}+ Q_\mathrm{joule}+ Q_\mathrm{visc}, \\
&\frac{\partial \mathbf{B}}{\partial t} = \nabla \times (\mathbf{v}\times \mathbf{B})+ \eta \nabla^2 \mathbf{B},\\
&\epsilon  =\frac{P}{(\gamma -1)\rho},
\end{align}
where $\rho$, $\mathbf{v}$, $\mathbf{B}$, and P are density, velocity vector, magnetic field vector, and gas pressure. Gravity is included. 
We use constant explicit resistivity ($\eta$) such that the magnetic Reynolds number is $R_m$=100.
We assume perfect gas specific heat of $\gamma=5/3$. Viscous heating $Q_\mathrm{visc}$ and Joule dissipation $Q_\mathrm{joule}$ are also included.

The dimension constants we use are, density $\rho_\mathrm{c}=1.67 \times 10^{-7}\ \mathrm{g}\ \mathrm{cm}^{-3}$, temperature $T_\mathrm{c}=5100~\mathrm{K}$, pressure $P_\mathrm{c}=7.16\times 10^3\ \mathrm{erg}\ \mathrm{cm}^{-3}$, length $H_\mathrm{c}=180 \ \mathrm{km}$ and magnetic field strength $B_\mathrm{c}=300 \ \mathrm{G}$. From the above units we get velocity $v_\mathrm{0}=2.1\ \mathrm{km} \ \mathrm{s}^{-1}$ and time $t_\mathrm{0}=85.7\ \mathrm{s}$

The numerical box consists of an adiabatically stratified subphotospheric layer at $-3.6\ \mathrm{Mm}\le z < 0 \ \mathrm{Mm}$, an isothermal photospheric-chromospheric layer at $0 \ \mathrm{Mm} \le z < 1.9 \ \mathrm{Mm} $, a transition region  at $1.9 \ \mathrm{Mm} \le z < 3.2 \ \mathrm{Mm}$ and an isothermal coronal at $3.2 \ \mathrm{Mm} \le z < 25.2 \ \mathrm{Mm}$.

We assume to have a field-free atmosphere in hydrostatic equilibrium. The initial distribution of temperature (T), density ($\rho$), gas  ($P_\mathrm{g}$) and magnetic ($P_\mathrm{m}$) pressure is shown in Fig.  \ref{fig:stratification_instability}a.

For the initial magnetic field, we assume a horizontal cylindrical non-twisted magnetic flux tube located at $2.1 \ \mathrm{Mm}$ below the photosphere. The axis of the flux tube is oriented along the $y$-direction, so the transverse direction is along $x$ and height is in the $z$-direction.

The axial field of the tube is
\begin{equation}
B_{y}=B_\mathrm{0} \exp(-r^2/R^2),
\end{equation}
where $R=450$~km is a measure of the tube's radius and $r$ the radial distance from the tube axis. In order to make the flux tube to emerge, we assume a density distribution similar to the work by \citet{Archontis_etal2004}, which makes the middle part of the flux tube less dense and thus buoyant. The density deficit is
\begin{equation}
\Delta \rho = \frac{p_\mathrm{t}(r)}{p(z)} \rho(z) \exp(-y^2/\lambda^2),
\label{eq:deficit}
\end{equation}
where $p$ is the external pressure and  $p_\mathrm{t}$ is the pressure within the flux tube (magnetic plus gas). The parameter $\lambda$ is the length scale of the buoyant part of the flux tube. 
In our numerical experiments, we use $\lambda=5 \ (0.9 \ \mathrm{Mm})$ and $\lambda=10 \ (1.8 \ \mathrm{Mm})$. 
The field strength of the flux tube is $B_\mathrm{0}=2780 \ \mathrm{G}$.

The computational box has a size of $28.8\times28.8\times28.8 \ \mathrm{Mm}$ in the $x$,$y$,$z$ directions, in a $457\times457\times457$ grid. 
The resolution is 
such that, in the sites with steep field gradients (e.g. current layers), the magnetic Reynolds number is (on average) larger 
than the effective Reynolds number. This indicates that the 
explicit resistivity dominates numerical effects. We assume open boundary conditions in the $x$ direction, periodic boundary conditions $y$ direction, and a wave damping zone for $z>$~26 Mm, which decays exponentially the velocity, gas pressure and density fluctuations.

\begin{figure}
\centering
\includegraphics[width=0.4\textwidth]{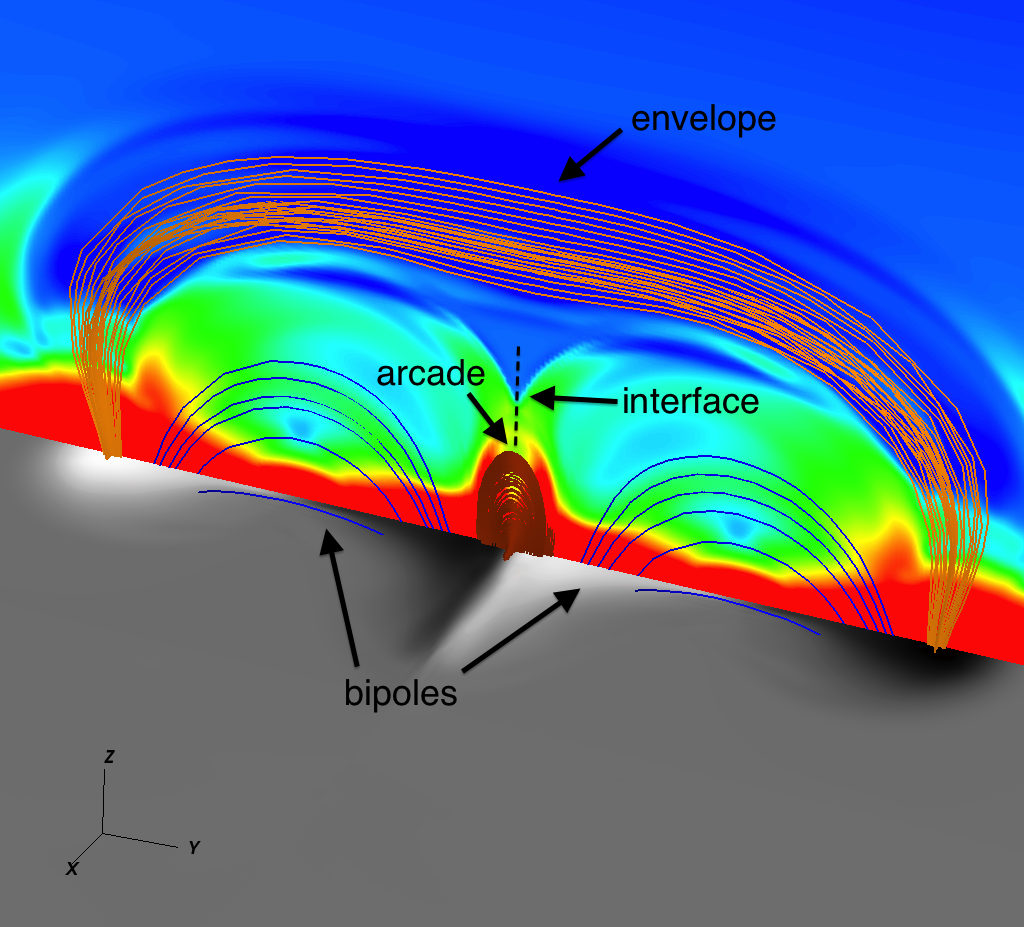}
\caption{\textit{Grayscale}: $B_{z}$ component of the magnetic field at the mid-photosphere ($z=0.9$~Mm), ranging 
from $-40$~G to $40$~G. \textit{Colored scale}: density at $yz$-midplane, ranging 
from 1.7$\times$10$^{-14}$~g~cm$^{-3}$ (red) to 1.7$\times$10$^{-16}$~g~cm$^{-3}$ (blue). The lines correspond 
to fieldlines connecting the bipoles (blue), fieldlines of the arcade (red) and fieldlines of the envelope field (orange). Time is $t=250$~min.}
\label{fig:loop_arcade}
\end{figure}
\section{Case of $\lambda=10$}
\label{sec:lambda_10}


\begin{figure*}
\centering
	\includegraphics[width=\textwidth]{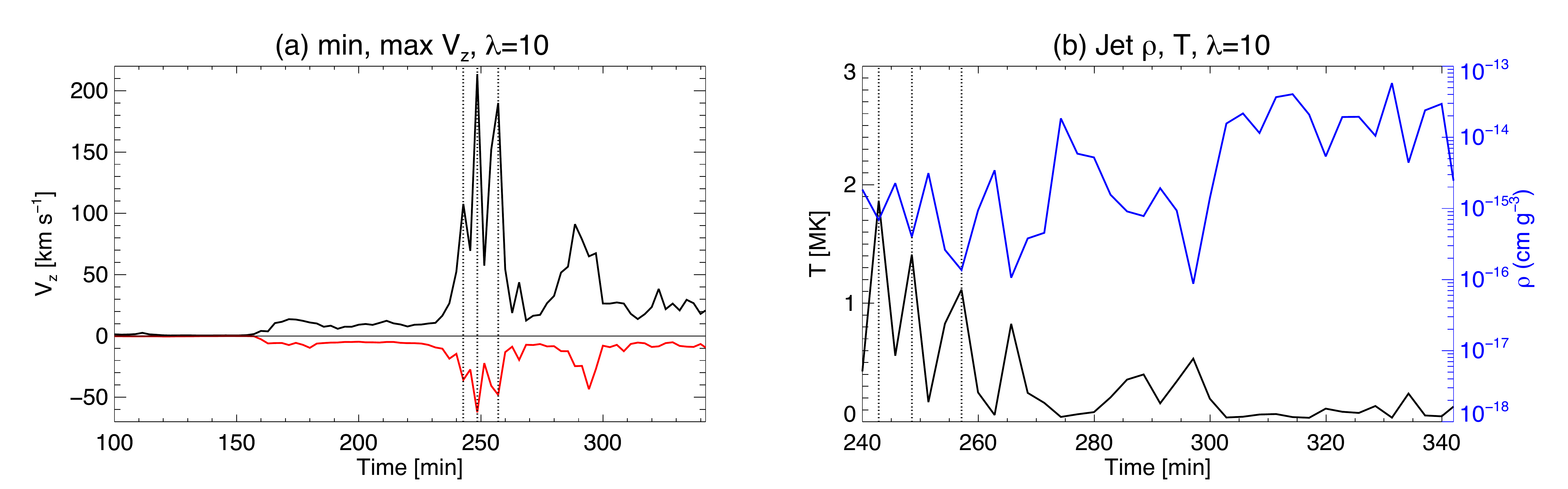}\;
	\caption{\textbf{(a)}: Temporal evolution of maximum (black) and minimum (red) $V_z$ above the photosphere. Vertical dashed lines correspond 
to the emission of the bi-directional jets at $t=242,248, 257$~min. \textbf{(b)}: Maximum temperature (black) and density (blue) around the locations of 
maximum $V_z$.
	}
	\label{fig:vz_temp_rho_10}
\end{figure*}


In this section we describe the  $\lambda=10$ case. 
We study the emergence of the flux tube through the highly stratified atmosphere and the onset of the associated dynamic phenomena in the EFR. 

\subsection{Initial emergence of the flux tube}

As the magnetic field emerges from the solar interior, the flux tube reaches the photosphere where it becomes compressed. In order for the flux tube to emerge above the photosphere, it has to satisfy the buoyancy instability criterion \citep{Acheson1979, Archontis_etal2004}
\begin{equation}
-H_\mathrm{p} \frac{\partial}{\partial z}(\log B) > -\frac{\gamma}{2}\beta\delta +k^2_{\parallel} \left( 1+ \frac{k^2_{z}}{k_{\perp}^2}\right),
\label{eq:instability_criterion}
\end{equation}
where $H_\mathrm{p}$ is the pressure scale-height, $B$ is the magnitude of the magnetic field, $\gamma=5/3$ is the specific heats ratio, $\beta$ is the plasma parameter, $\delta=-0.4$ is the superadiabatic excess and $k$ are the perturbation wavenumbers, where $k_\parallel$ and $k_\perp$ are the horizontal components parallel and vertical to the magnetic field, and $k_{z}$ is the vertical component.
To calculate these wavenumbers, we measure the size of the flux tube sections below the photosphere. 
We find $l_{x}\approx0.5$~Mm and $l_{y}\approx1.4\ \mathrm{Mm}$ and approximately the radius of the tube as the vertical width $l_{z}\approx0.45$~Mm. These estimates are taken at $t=154$~min. The corresponding wavevector components are $k_\parallel\approx4.5\times10^{-8} \ \mathrm{cm}^{-1}$, $k_\perp\approx1.2\times10^{-7} \ \mathrm{cm}^{-1}$ and $k_{z}\approx1.4\times10^{-7} \ \mathrm{cm}^{-1}$. Since the wavevector term in the instability criterion is small ($4.6\times10^{-15} \ \mathrm{cm}^{-2}$) in comparison to the $\beta$ term (which has minimum values of 10$^{-12}$), we neglect it, and after dividing the inequality with its right part, we get approximately:
$-H_\mathrm{p} \frac{\partial}{\partial z}(\log B) / ( -\frac{\gamma}{2}\beta\delta) > 1$.
We refer to the left side of this equation as the ``instability term''.

At the center (around $x=0,\,y=0,\,z=0$) of the EFR, the field magnitude decreases as the flux tube expands. At the areas sideways of the center, along the $y$-direction,  the field is compressed, leading to a smaller gradient of the magnetic field and a lower $\beta$. This makes the instability term greater than 1 in these areas, while it remains below unity around the center of the EFR. Thus, the flux tube does not emerge at the center of the EFR, but it forms two side bipoles, as shown in Fig. \ref{fig:magneto_tail}a.
In Fig. \ref{fig:stratification_instability}b, \ref{fig:stratification_instability}c we plot the temporal evolution of the instability term at two positions in the numerical box. In Fig. \ref{fig:stratification_instability}b, we calculate the instability term at the center of the EFR. We confirm that the field does not emerge at this region (Fig. \ref{fig:magneto_tail}a). The flux tube becomes unstable at two areas along the flux tube axis, at $(x,y)=(0,\pm4.8$~Mm). There, the instability term becomes greater than 1 (Fig. \ref{fig:stratification_instability}c, calculated at $y=4.8$~Mm). This leads to the emergence of two bipolar regions.

At $t=169$~min (Fig. \ref{fig:magneto_tail}a), the two bipoles have emerged into the photosphere. Hereafter, we refer to polarities A,D as outer polarities, and C,B as inner polarities. During the evolution, the polarities at each bipoles expand along the y-direction. At the photosphere, the fieldlines that join the two polarities of each bipole are almost horizontal due to the absence of twist. If the flux tube was twisted, we would have detected magnetic ``tails'' on the two sides of the PIL in each bipole \citep[e.g.][]{LopezFuentes_etal2000,Chandra_etal2009}. As the magnetic field rises above the photosphere, the magnetized volume of each bipole expands and the attached magnetic field lines form two magnetic bipoles (Fig. \ref{fig:magneto_tail}b).

\begin{figure*}
 \centering
 \subfloat[]{\label{fig:fieldlines_81}\includegraphics[width=0.35\textwidth]{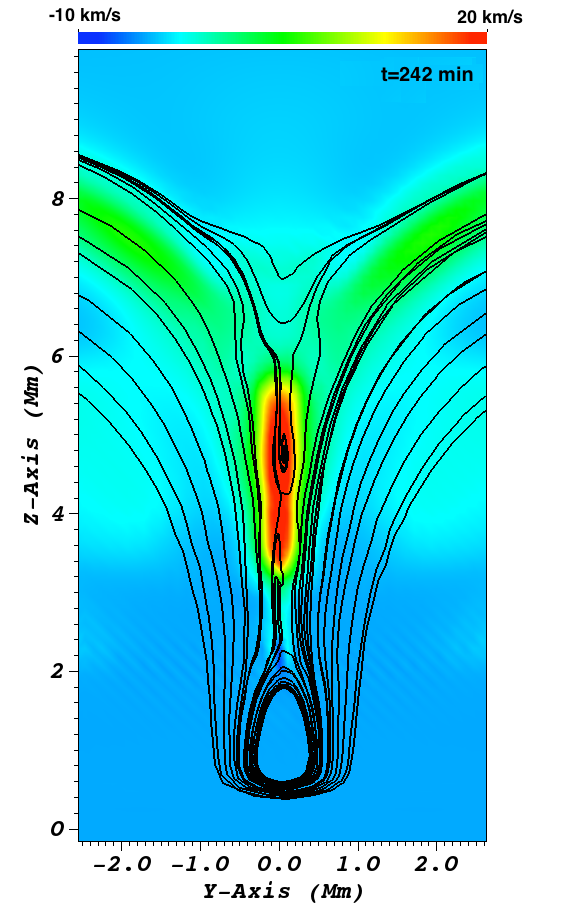}}\;
 \subfloat[]{\label{fig:fieldlines_93}\includegraphics[width=0.35\textwidth]{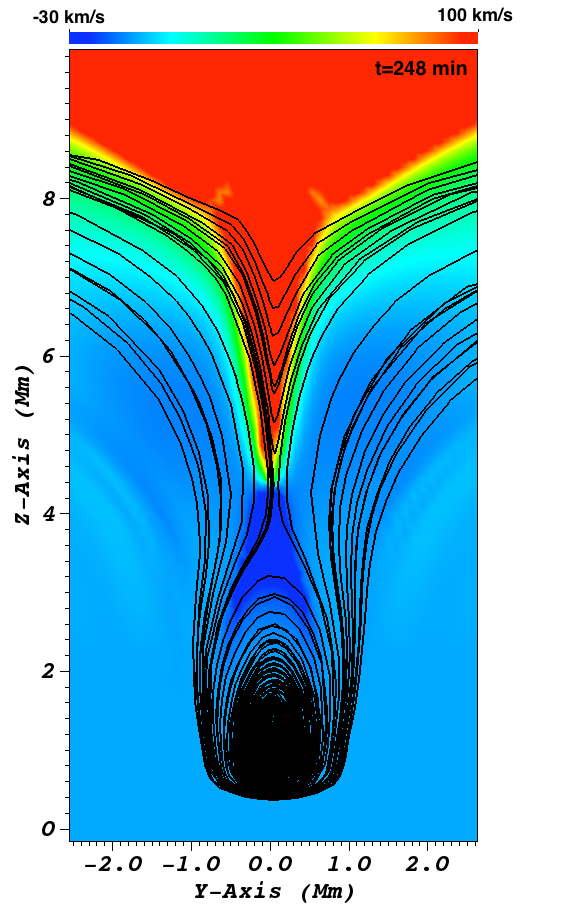}}\;
 \caption{The $V_z$ component of the velocity field on the $yz$-midplane. Field lines show the projection of the full magnetic field vector onto that plane. \textbf{(a):} Prior to plasmoid ejection at $t=242$~min. \textbf{(b):} Jet ejection at $t=248$~min. }
  \label{fig:plasmoid_10}
\end{figure*}
\begin{figure*}
 \centering
 \includegraphics[width=\textwidth]{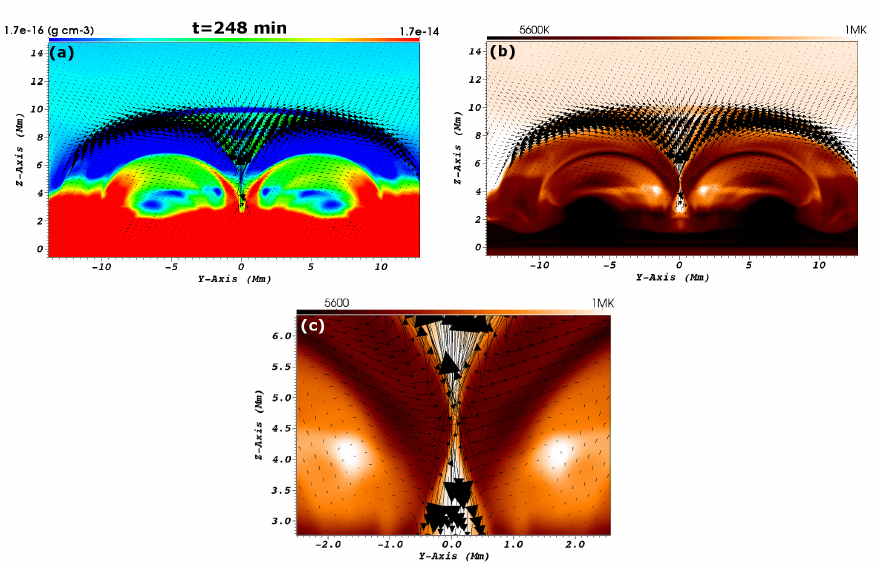}
 \caption{Density (a) and temperature (b, c) distribution at the vertical $yz$-midplane. Overplotted is the projection of 
the velocity field vector onto the plane. Bottom: A close-up of the reconnection site.
 }
 \label{fig:jets}
\end{figure*}

\subsection{Interaction of the magnetic bipoles}
In the following, we study the effect of the interaction of the magnetic bipoles on the solar atmosphere. Firstly, we focus on the formation of plasmoids and jets, and secondly on the transfer of plasma and flux into the solar atmosphere.

\subsubsection{Jets and plasma transfer}
\label{sec:jets_plasma_tranfer}

After $t=171$~min, the two magnetic bipoles come into contact and a current layer is formed at their interface. Eventually, the long and thin current layer becomes subject to the tearing mode instability. 
Thus, plasmoids are formed within the interface and later on they are ejected out of it. This induces inflows towards 
the interface and initiates effective reconnection of the oppositely directed fieldlines 
of the bipoles into contact. The dynamics of plasmoids in a similar configuration (i.e. interaction between two individual twisted flux tubes) was studied 
by \citet{Archontis_etal2007}.
Fig. \ref{fig:loop_arcade} shows the overall topology 
of the fieldlines at a later stage of the evolution, when two new magnetic flux systems (apart from the two magnetic bipoles) have been developed.
More precisely, reconnected fieldlines join the two outer polarities forming the envelope field 
(Fig. \ref{fig:loop_arcade}, orange lines) and reconnected fieldlines join the inner polarities forming an arcade-like structure (Fig. \ref{fig:loop_arcade}, red lines).
Because of the emergence of the field and the gravitational draining of the plasma, low-atmosphere material accumulates at the interface, filling the area with cool and dense plasma (Fig. \ref{fig:loop_arcade}, green color around interface in the $yz$-plane).

In Fig. \ref{fig:vz_temp_rho_10}a, we plot the temporal evolution of maximum (black) and minimum (red) $V_z$ above the photosphere. 
Up to $t\approx240$~min we find moderate upflows/downflows ($5-10$~\kms) due to the emergence of the two bipoles and 
the draining of the plasma at the periphery of the magnetic bipoles respectively.
The first major upflow occurs at $t=242$~min (marked with a vertical dashed line). This is associated with the ejection of a plasmoid from the 
interface current layer into the corona. The ejection is driven by the tension of the reconnected fieldlines at the interface. The ejected 
plasmoid is cool ($\approx 10^{5} K$) and dense (1-2 orders of magnitude heavier than the background plasma in the corona). During the evolution 
of the system, several plasmoids are formed 
and eventually being ejected from the 3D interface. The ejection of the plasmoids is followed by (fast) reconnection of fieldlines at the interface 
and the emission of a series of bi-directional flows (jets). The occurence of the bi-directional jets is shown in Fig. \ref{fig:vz_temp_rho_10}a for 
$t \geq 242$~min.

Fig.~\ref{fig:vz_temp_rho_10}b shows the maximum temperature and density around the locations of maximum $V_z$. We find that the 
temperature evolution of the temperature is also highly impulsive. 
In fact, there is a very good temporal                                                                          
correlation between the bi-directional flows and the sudden temperature enhancements, which is an indication of a reconnection-driven plasma acceleration at the interface. The impulsive behavior shows that there are many jets, which are emitted
from the interface, during the interaction of the magnetic bipoles. The first three high-speed jets occur in the
period $t\approx 240-260$~min. They have temperatures in the range $1-1.5$~MK and densities between $10^{-16}-10^{-15} \ \mathrm{g} \ \mathrm{cm}^{-3}$  (Fig.~\ref{fig:vz_temp_rho_10}b)
and, thus, they might account for soft X-ray jets.
The following jets (for $t\geq 260$~min) are (in general) less energetic, with velocities up to $\approx50$~\kms (with an exception, at
$t\approx 295$~min, where $V_z \approx 100$~\kms) and temperatures lower than $1$~MK.
The general decay of the onset of high-speed jets is due to the gradual decrease in the available
flux and energy of the system that drives the ejections. The available energy comes only from the emerging magnetic field. Thus, this energy                 
is limited and it unavoidably becomes exhausted.

Fig.~\ref{fig:plasmoid_10}a is a close-up of the interface showing the upward motion of the plasmoid before the onset of the first bi-directional jet.
The colorscale shows the vertical component of the velocity field and the overplotted fieldlines (black lines)
show the projected magnetic field vector onto the plane. The fieldlines reveal the existence of the plasmoid 
(magnetic ``island'' in the 2D plane) at the interface.
The upflow (red color) illustrates the upward motion (ejection) of the plasmoid at $t=242$~min. The topology of the fieldlines show that the ejection is indeed
driven by the tension of the V-shaped reconnected fieldlines underneath the plasmoid. The downward product of reconnection is the emission of a weaker
downflow (blue color) and the formation of the O-shaped arcade.

Fig.~\ref{fig:plasmoid_10}b shows the emission of a strong bi-directional jet after the expulsion of the plasmoid out of the 
interface. Now, the fieldlines from the 
interacting bipoles come into direct contact (since the plasmoid in between them has been disposed) at the interface and reconnect. This is reminiscent 
of the plasmoid-induced reconnection scenario, suggested by \citet{Yokoyama_etal1994}. The V-like reconnected 
fieldlines, which are released upwards, accelerate 
the reconnection outflow (jet). The downward-released fieldlines are added to the arcade, which thus, undergoes an apparent rising motion and it 
grows in size.

Fig.~\ref{fig:jets}a-c show the plasma dynamics during the emission of the reconnection jet at $t=248$~min. The bi-directional flow starts at a coronal 
height ($z\approx 4.5 Mm$). The upward reconnection flow runs with a velocity of $190$~\kms and it carries relatively dense and hot ($\geq 1.0$~MK)
plasma into the corona (6a, 6b). It reaches a height of $\approx 10$~Mm, where it collides with the envelope field and it is diverted into 
two side flows, which run with velocities of $\approx68$~\kms. Figure.~\ref{fig:jets}c is a close-up of the reconnection region showing clearly the inflows 
towards the reconnection site, which are enhanced after the ejection of the plasmoid, and the profound hot, bi-directional jet driven by reconnection.

\subsubsection{Undulation of the PIL, fieldline topology and structure of the jets}
\label{sec:undulation_of_pil}
\begin{figure*}
\centering
\includegraphics[width=0.85\textwidth]{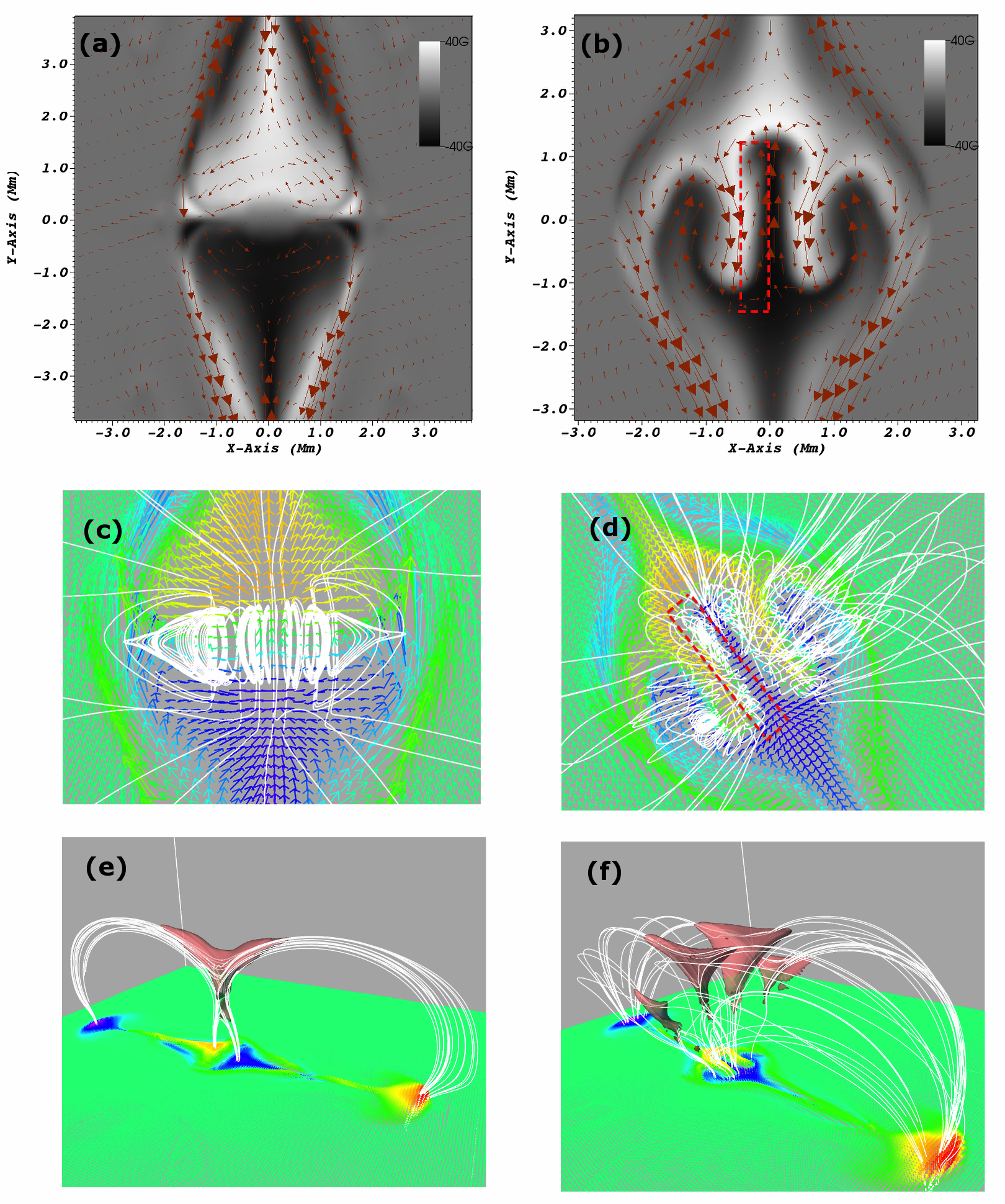}\;
\caption{\textbf{(a):} $B_z$ component of the magnetic field at the mid-photosphere ($z=0.9$~Mm), with the velocity 
field overplotted at $t=243$~min and \textbf{(b):} $t=286$~min. \textbf{(c):} Fieldlines traced around the center of 
the EFR at $t=243$~min, showing the arcade. Overplotted is the full magnetic field vector colored by the $B_z$ component 
of the field at $z=0.9$~Mm. Red is positive and blue is negative. The red rectangular inset marks the same region as the 
one in (b). \textbf{(d):} Fieldlines at $t=286$~min, showing the deformation of the PIL. \textbf{(e):} Isosurface 
of $V_z>60$~km~s$^{-1}$ showing the V-shaped jet at $t=248$~min. \textbf{(f):} The emission of several jets at $t=286$~min (isosurface of $V_z >60$~km~s$^{-1}$). }
\label{fig:mag10}
\end{figure*}

\begin{figure*}
\centering
	\includegraphics[width=\textwidth]{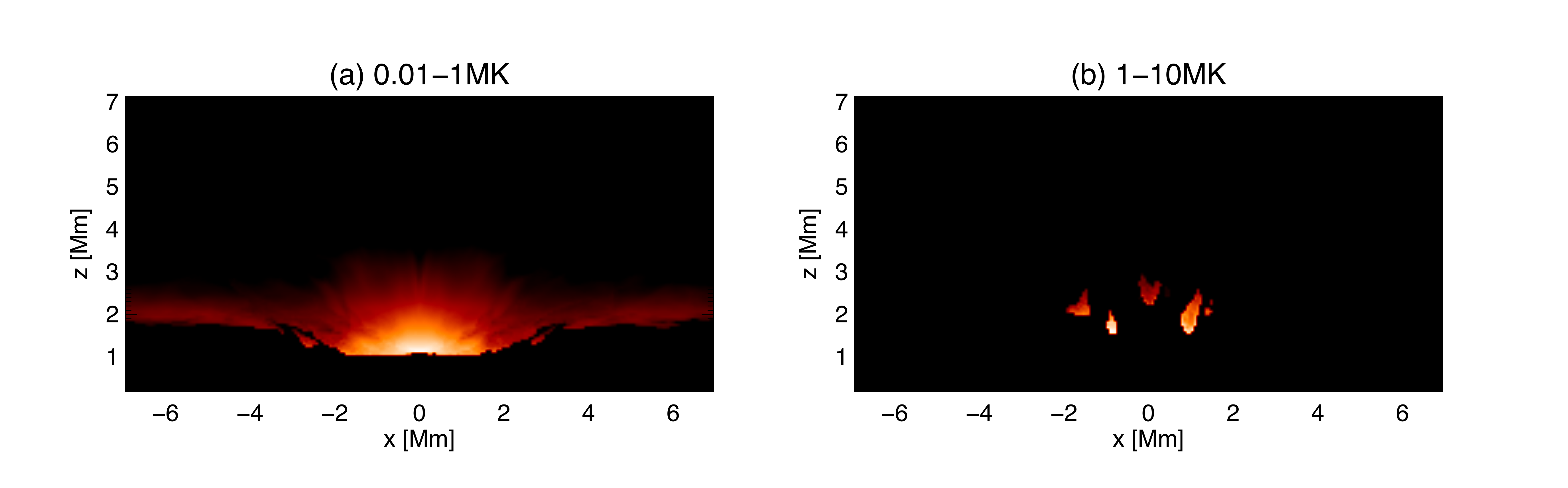}\;
	\caption{ Visualization of the heating term (H) at $t=268$~min for plasma having 
temperatures between \textbf{(a):} $0.01-1$~MK and \textbf{(b):} $1-10$~MK.
	}
	\label{fig:synthetic_images_10}
\end{figure*}

One of the interesting features in this experiment is the evolution of the arcade structure at the interface. To study this, we 
firstly show the temporal evolution of the $B_z$ distribution at the middle of the photosphere (Fig. \ref{fig:mag10}a, \ref{fig:mag10}b). 
Initially (Fig. \ref{fig:mag10}a, $t=243$~min), the inner polarities collide and the PIL in between them becomes 
almost perfectly aligned with the $x$-axis at the center 
of the emerging flux region. 
However, later we find that the PIL adopts an undulating shape (Fig. \ref{fig:mag10}b, $t=286$~min). The overplotted velocity field on this 
plane shows that the deformation of the PIL is associated with plasma motions along the $y$-axis, towards the PIL. 
A similar deformation was reported in the 
work by \citet{Archontis_etal2010}: the tails of the magnetic field on the two sides 
of the PIL suffered by converging flows (inflows), which were produced due to the 
rapid expansion of the field above the photosphere. 
Here, we find that a similar process is at work. The finger-like segments, which are developed across 
the PIL, are sites with low density and pressure. 
At this height ($z=0.9$~Mm), the magnetic field is also weak and the kinetic energy density is such 
that the magnetic field is advected towards the PIL.
The low pressure at various points is produced due to the rapid ejection of several dense plasmoids from along 
the interface. The plasma then moves towards the low-pressure areas (inflows), deforming the PIL.

Figure \ref{fig:mag10}c shows the 3D topology of the fieldlines around the arcade before the deformation ($t=243$~min). Overplotted is the full magnetic field vector at the photosphere, colored by the vertical ($B_z$) component (red is positive, blue is negative). 
The arcade consists of fieldlines that join the inner polarities all along the PIL. 
The fieldines have been traced from random positions within the two inner polarities. We note that the fieldlines of the arcade towards its two ends (along the $x$-direction) are more inclined than the fieldlines at the middle of the arcade. 
This is due to the fan-like shape of the interacting fieldlines of the two magnetic bipoles. We also note the cusp-like shape at the top of the arcade's fieldlines. This indicates the occurrence of reconnection above them. Left and right from the arcade, we find expanding fieldlines that belong to the interacting 
bipoles. Envelope fieldlines are not shown in this figure. 

Figure \ref{fig:mag10}d shows the topology of the fieldlines around the deformed PIL at $t=286$~min. 
We note that after the deformation, the magnetic fieldlines becomes 
highly twisted all along the PIL.
Some segments of the PIL are now oriented along the $y$-axis. For instance, the segment which is outlined by the red rectangular inset 
in Fig. \ref{fig:mag10}d, is the same with the segment around $x=-0.5$~Mm and between $y=-1$~Mm and $y=1$~Mm in Fig. \ref{fig:mag10}b.
Now, the relative motion of the inflows on the two sides of this segment induce a shearing 
(Fig. \ref{fig:mag10}b, motions towards negative $y$-axis and positive $y$-axis around $x=-0.3$~Mm, as the positive (negative) inner polarity moves towards the negative (positive) polarity).
The sheared and oppositely directed fieldlines across the PIL's segment reconnect, producing a small twisted flux rope and a reconnection jet. 
This process occurs at several (similar) segments along the PIL, providing an additional source of twist for the fieldlines around the PIL and triggering the onset 
of more jets at the interface. 
Thus, we find that the deformation of the PIL leads to the parallel emission of new jets from the interface. The highly intermittent evolution of $V_{z}$ for $t>280$~min in Fig. \ref{fig:vz_temp_rho_10}a is due to this effect. 

The 3D structure of the emitted jets before and after the deformation of the PIL is shown in Fig. \ref{fig:mag10}e, \ref{fig:mag10}f. 
In Fig. \ref{fig:mag10}e we show the 3D structure of the $t=248$~min jet, by visualizing 
the total velocity field as an isosurface (red color) with $V > 60\ \mathrm{km\,s^{-1}}$. 
At low heights, the reconnection upflow has the shape of a sharp wedge with its length oriented 
vertically, above the reconnection site. 
At larger heights, it becomes confined by the envelope field and becomes diverted into 
two lateral flows, following the reconnected fieldlines. Thus, the overal configuration of the upward jet adopts a V-like shape. 
The magnetic fieldlines (white), which have been traced from within the lateral parts of the jet illustrate the envelope field. 
We also identify the two magnetic bipoles and the arcade underneath the reconnection site. At this time of the evolution, the downward reconnection jet is much weaker and, thus, it appears as a horizontal ribbon-like structure just above the arcade. Overplotted is the full magnetic field vector (horizontal cut) at $z= 0.9\,\mathrm{Mm}$. 

In Fig. \ref{fig:mag10}f, we visualize the total velocity field as an isosurface with $V>60\,\mathrm{km\,s^{-1}}$ at $t=311$~min. 
The striking difference  
is the appearance of several jets (here, four jets) from the interface. 
All the upward jets have similar V-like shapes. The parallel emission of several hot jets from the interface starts after the 
deformation of the PIL and it is followed by intermittent heating of the plasma at the interface.
The latter is discussed more in the following section.

\begin{figure*}
 \centering
 \includegraphics[width=0.9\textwidth]{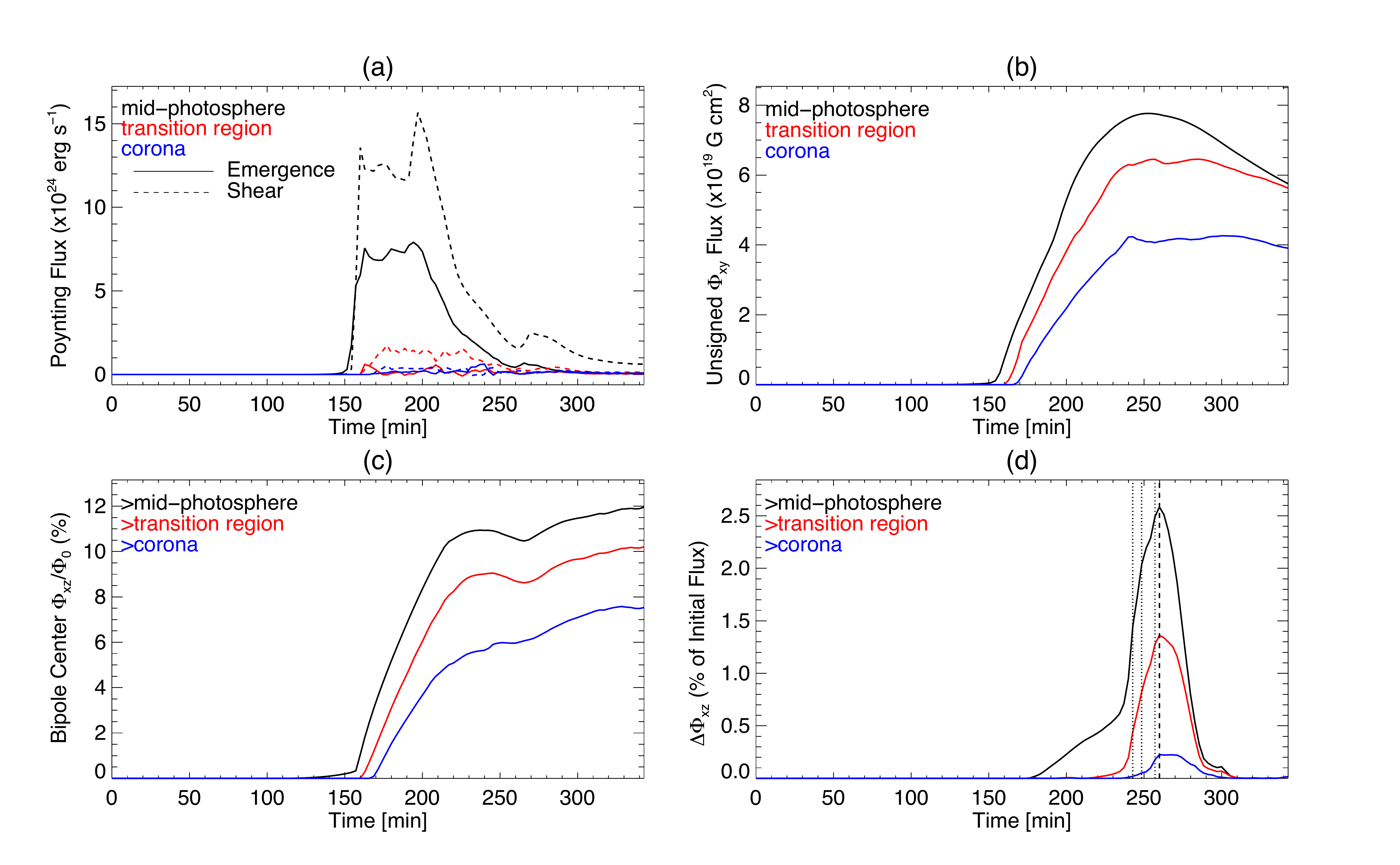}
 \caption{ \textbf{(a):} Temporal evolution of the Poynting Flux. Solid lines show the emergence term,  $F_{\mathbf{emergence}}$, and dashed lines show the shearing term, $F_{\mathbf{shear}}$. \textbf{(b):} The unsigned magnetic flux $\Phi_{xy}$. \textbf{(c):} The axial flux for the bipole's center in the positive $y$-direction (bipole AB). \textbf{(d):} Flux difference $\Delta\Phi_{xz}$. The dashed vertical line marks the initiation of the undulation of the PIL and the dotted vertical lines show the times for the onset of the hight-speed jets.}
 \label{fig:fluxes_10}
\end{figure*}

\subsubsection{Intermittent heating at the solar atmosphere.}
\label{sec:synthetic_observation}

To study how much and where the dense plasma is heated during the emission of the jets, we calculate the term:
\begin{equation}
H=\int_{T_{min}}^{T_{max}}\rho^2 dy\,,
\end{equation}

We do the calculation for temperatures in the range of $0.01-1$~MK and $1-10$~MK, as a preliminary attempt to separate the cooler (i.e. $< 1$~MK) from the 
hotter component of the plasma (i.e. $>1$~MK). Hereafter, for simplicity, we call this term the ``heating term''. Figure~\ref{fig:synthetic_images_10} shows the heating term after 
the emission of the first four successive jets (i.e. at $t=268$~min), during which the temperature enhancement of the associated plasma is more pronounced 
(see Fig.~\ref{fig:vz_temp_rho_10}b).

In the lower temperature regime (Fig.~\ref{fig:synthetic_images_10}a), we find that heating is more evident in a fan-like arcade, which extends (vertically) from the chromosphere to the low corona. 
The arcade is formed by the reconnected fieldlines, which are released downwards, after the onset of the jets at the interface. The fan-like shape is formed 
naturally due to the 3D lateral expansion of the interacting magnetic bipoles and it is in agreement with the 3D shape of the arcade shown in Fig.~\ref{fig:mag10}c. 

In the higher temperature regime (Fig.~\ref{fig:synthetic_images_10}a), the heating is not smoothly distributed across the arcade, but it is much more intermittent. There are four small scale 
($< 1$~Mm) brightenings, which all occur within the area between the chromosphere and the transition region. These brightenings occur exactly under the 
reconnection sites of the four bi-directional jets, which were emitted during the $t=240-260$min period. The reconnection, which triggers the jets, occurs 
mainly in the corona. However, the downward jet(s) collide with the top of the arcade forming a termination shock. There, the plasma is compressed locally 
at high temperatures, causing heating of the dense material. 

A very similar process, regarding the intermittent heating, was reported in recent more 
realistic simulations (radiative transfer, MHD) numerical experiments \citep{Archontis_etal2014}. In those simulations, small-scale flaring activity was produced at the current sheets between 
interacting magnetic loops. The small flares were triggered by plasmoid-induced reconnection at the interface current layers. In the present work, we are 
witnessing an analogous process: patchy reconnection associated with the ejections of plasmoids leads to plasma heating, which resembles the onset of nano-micro 
flares in the model by \citet{Archontis_etal2014}. 
The present model does not include all the necessary energy terms (radiative transfer, heat conduction, 
radiation, etc.) for a complete treatment of the thermodynamical behavior of the system. The lack of these terms affects the estimated value of temperature (e.g. of the jets) in the present experiments.

\subsubsection{Energy transfer and magnetic flux}
\label{sec:energy_transfer}

To study the transfer of flux into the solar surface and above, we first calculate the Poynting flux.
We split it into the ``emergence term'', which is the part of flux representing the direct emergence of the magnetic 
field, and the ``shear term'', which represents the flux due to the horizontal motions \citep{Kusano_etal2002,Magara_etal2003}:
\begin{align}
&F_{\mathbf{emergence}}=\frac{1}{4\pi}\int_{z_0} (B_x^2+B_y^2)v_zdxdy, \label{eq:emergence}\\
&F_{\mathbf{shear}}=-\frac{1}{4\pi}\int_{z_0} (B_x v_x+B_y v_y)B_zdxdy. \label{eq:shearing}
\end{align}
We calculate these fluxes at the mid-photosphere (${z_0}=0.8$~Mm), the base of the transition region (${z_0}=2$~Mm) and the low 
corona (${z_0}=3.2$~Mm).

In Fig. \ref{fig:fluxes_10}a,  we plot the emergence term (solid line) and the shearing term (dashed line).
At the mid-photosphere (black), we find the emergence term to be greater than the shearing term 
between $t=140-155$~min, as expected during the beginning of the emergence. The shearing term 
becomes greater than the emergence term, when the inner polarities collide. Owing to this collision, plasma is 
pushed sideways along the PIL and towards the $x$-direction, contributing to the 
shearing term (e.g. Fig. \ref{fig:mag10}a, flows along the PIL).
Both terms reach their maximum value at $t\approx200$~min and soon thereafter start to decrease. The drop in the two terms is due to the fact that the dynamical emergence and expansion of the field becomes less pronounced at the later stage of the evolution of the system.
At larger heights (blue lines), we find that the shearing term is more dominant due to the expansion of the field.

To study the vertical flux transfered at the same heights, we calculate the unsigned magnetic flux through the $xy$-plane,
\begin{equation}
\Phi_{xy,z_0}=\int_{z_0} | B_z | dxdy.
\end{equation}
At the photosphere, $\Phi_{xy}$ increases monotonically until $t=255$~min (Fig. \ref{fig:fluxes_10}b, black line). The drop of the flux thereafter is evidence of flux cancellation at the photosphere, even during the deformation of the PIL after $t=260$~min.
Measuring $\Phi_{xy}$ at larger heights (red and blue lines) shows that less vertical flux is transported higher, as expected.

\begin{figure*}
 \centering
 \includegraphics[width=0.9\textwidth]{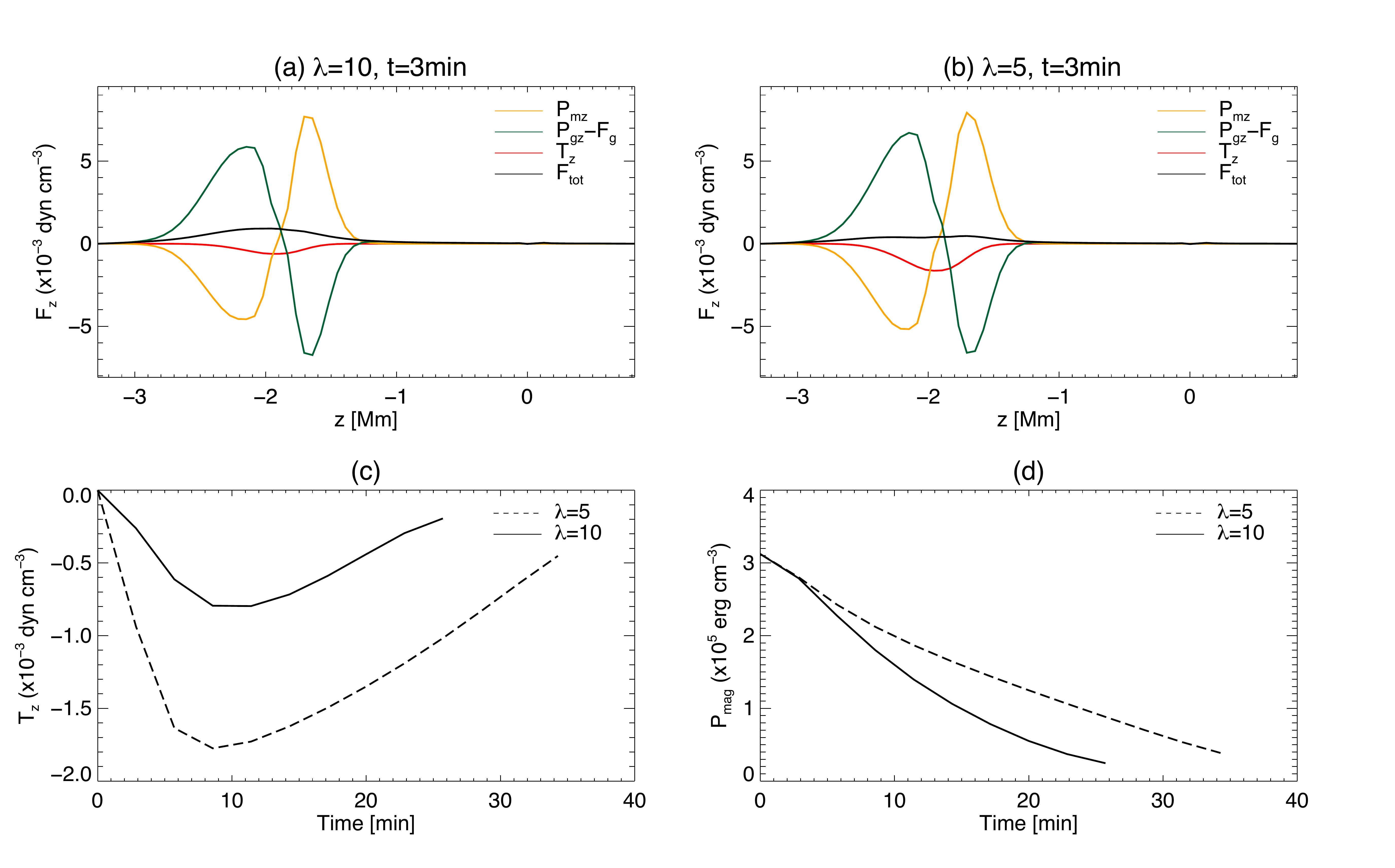}
 \caption{First row: $z$ component of the forces along the vertical direction ($x=0$, $y=0$) at $t=3$~min, for \textbf{(a)} $\lambda=10$  and \textbf{(b)} $\lambda=5$. Second row: Temporal evolution of \textbf{(c)} $z$-component of the magnetic tension and \textbf{(d)} magnetic pressure, measured at the center of the flux tubes. We plot the quantities until the flux tubes reach the photosphere. 
 } 
 \label{fig:tube_center}
\end{figure*}

We also calculate the normalized axial flux through the $xz$-plane at a position $y_0$, from the heights ($z_0$) 
up to the top of the numerical box ($z_\mathrm{max}$), i.e.
\begin{equation}
\Phi_{xz,y_0}=\int_{z_0}^{z_\mathrm{max}}  B_y  dxdz \ / \int_{0}^{z_\mathrm{max}}   B_y (t=0)  dxdz.
\end{equation}
We calculate this axial flux at $y_0=\pm6$~Mm, which are located around the center of the bipoles. We find 
the same flux at both locations, meaning that the same amount of axial flux emerges at both bipoles. 
In Fig. \ref{fig:fluxes_10}c, we plot this flux for the bipole center in the positive $y$-direction. An 
interesting result is that for each bipole, only 12\% of the initial axial flux emerges above the 
photosphere (black line). From this, only 7.5\% gets above the corona (blue line). Adding the 
axial flux of the two bipoles, a sum of 24\% of the initial flux emerges above the photosphere and 15\% above the corona. Thus, this 
study shows that non-twisted subphotospheric fields with these characteristics are not capable of transfering a 
considerable amount of flux above the photosphere.

Owing to the non-twisted magnetic field in our initial conditions, there is only positive $B_y$ at the beginning of the simulation. The emergence and expansion of the field forms the $B_x$ and $B_z$ components. In our configuration, negative $B_y$ component at the $xz$-midplane could be formed through reconnection. This component is part of the arcade's magnetic field lines. In order to estimate how much negative $B_y$ is created on this plane, we calculate the normalized flux difference:
\begin{equation}
\Delta\Phi_{xz}=\frac{1}{2}\int_{z_0}^{z_\mathrm{max}} \left(  |B_y|   -  B_y\right) dxdz  \ / \int_{0}^{z_\mathrm{max}} B_y(t=0) dxdz.
\label{eq:flux_difference}
\end{equation}
Here $\Delta\Phi_{xz}$ measures only the flux of negative $B_y$ through the $xz$-midplane, from a height $z_0$ up to the top of the numerical box $z_\mathrm{max}$, normalized by the initial positive flux before the emergence. So, $\Delta\Phi_{xz}$ indicates how much of the initially positive flux is transformed into negative flux, through reconnection. This calculation is valid only at the $xz$-midplane and due to the absence of twist.

Figure \ref{fig:fluxes_10}d shows above the mid-photoshere (black line), flux difference increases after $t=170$~min. This is due to reconnection at the interface. This leads to the formation of new fieldlines that point in the negative $y$-direction, joining the two inner polarities, and forming the arcade. We find that $\Delta\Phi_{xz}$ increases rapidly after the major ejections (dotted vertical lines), peaking after the $t=240-260$~min series of jets. 
This maximum shows that 2.6\% of the initial $B_y$ flux of the system is transformed into negative $B_y$. 
In the corona (blue line), we only find 0.3\% of the flux. As a result, most of the post-reconnection flux at the interface resides between the photosphere and the low corona.

After $t=260$~min (vertical dashed line), we find that $\Delta\Phi_{xz}$ starts to decrease drastically. This is when the deformation of the PIL starts to occur (Fig. \ref{fig:mag10}b, \ref{fig:mag10}d). 
At this moment, the interface cannot be studied by using the 2D vertical $xz$-midplane, and $\Delta\Phi_{xz}$ does not represent an accurate measurement of the normalized flux difference.
This happens because, as we described previously, the field lines of the arcade follow the deformation of the PIL (Fig. \ref{fig:mag10}d). Thus,  some of the field lines which were previously perpendicular to the interface, they are now pointing towards the $x$-direction due to the undulation (e.g. Fig. \ref{fig:mag10}b, lines around $x=\pm0.3$~Mm would tend to be parallel to the $x$-axis). As a result, their $B_y$  ($B_x$) component decreases (increases), and consequently, $\Delta\Phi_{xz}$ decreases. Also, the field lines that do not rotate, tend to move away from the $xz$-plane due to the flows in the $y$-direction (e.g. around the EFR center, PIL position is located $x=0$, $y\approx 1.3$~Mm). This further decreases $B_y$ on the $xz$-midplane, and contributes to the decrease in $\Delta\Phi_{xz}$. 

\section{Case of $\lambda=5$ and comparison with $\lambda=10$}
\label{sec:lambda_5}


In the following, we describe the emergence of the non-twisted flux tube with $\lambda=5$, and make a comparison to the $\lambda=10$ case.

\subsection{Emergence from the solar interior to the photosphere}
\label{sec:emergence _from_interior}
\begin{figure*}
 \centering
 \includegraphics[width=0.9\textwidth]{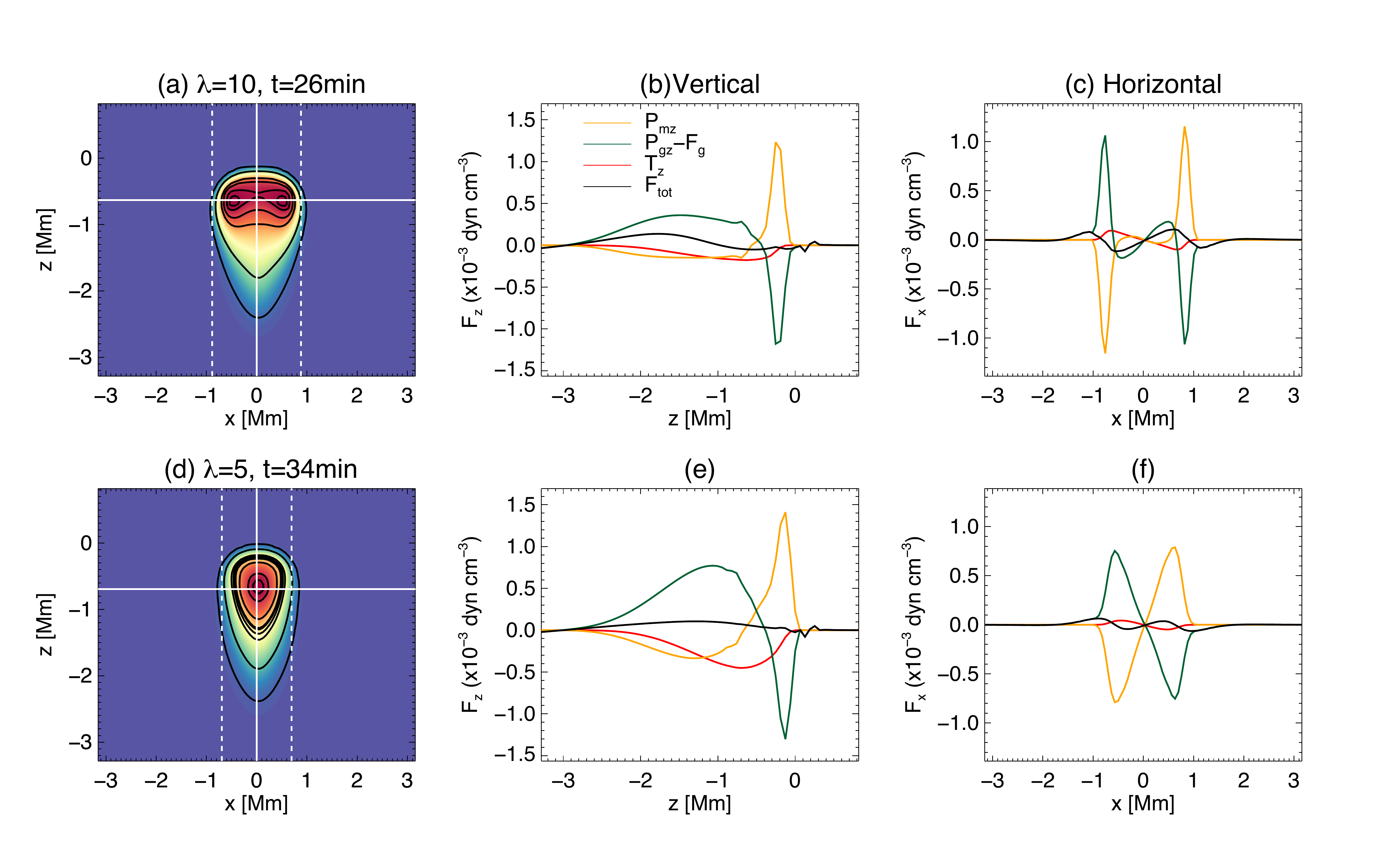}\;
 \caption{ Forces and flux tube cross sections of $\lambda=10$ and $\lambda=5$, before they reach the photosphere (at $t=26$~min and at $t=34$~min respectively).
 \textbf{(a,d):} $P_{mag}$ of the flux tubes at the $y=0$ cross section. The solid white lines (horizontal and vertical) pass through the flux tubes centers. Black $P_{mag}$ contours have the same values in both panels. Vertical white dashed lines mark the position of the flux tube's outer part. \textbf{(b,e):} $z$-component of forces along the vertical direction through the flux tube's center (direction marked with the vertical solid white line in (a,d) ). Green is gas pressure force minus gravity force, orange is magnetic pressure force, red is magnetic tension force and black the total force.  \textbf{(c,f)}: The $x$-component of forces along the horizontal direction through the flux tube's center (direction marked with the horizontal solid white line in (a,d) ).}
 \label{fig:interior}
\end{figure*}

\begin{figure*}
 \centering
 \includegraphics[width=0.85\textwidth]{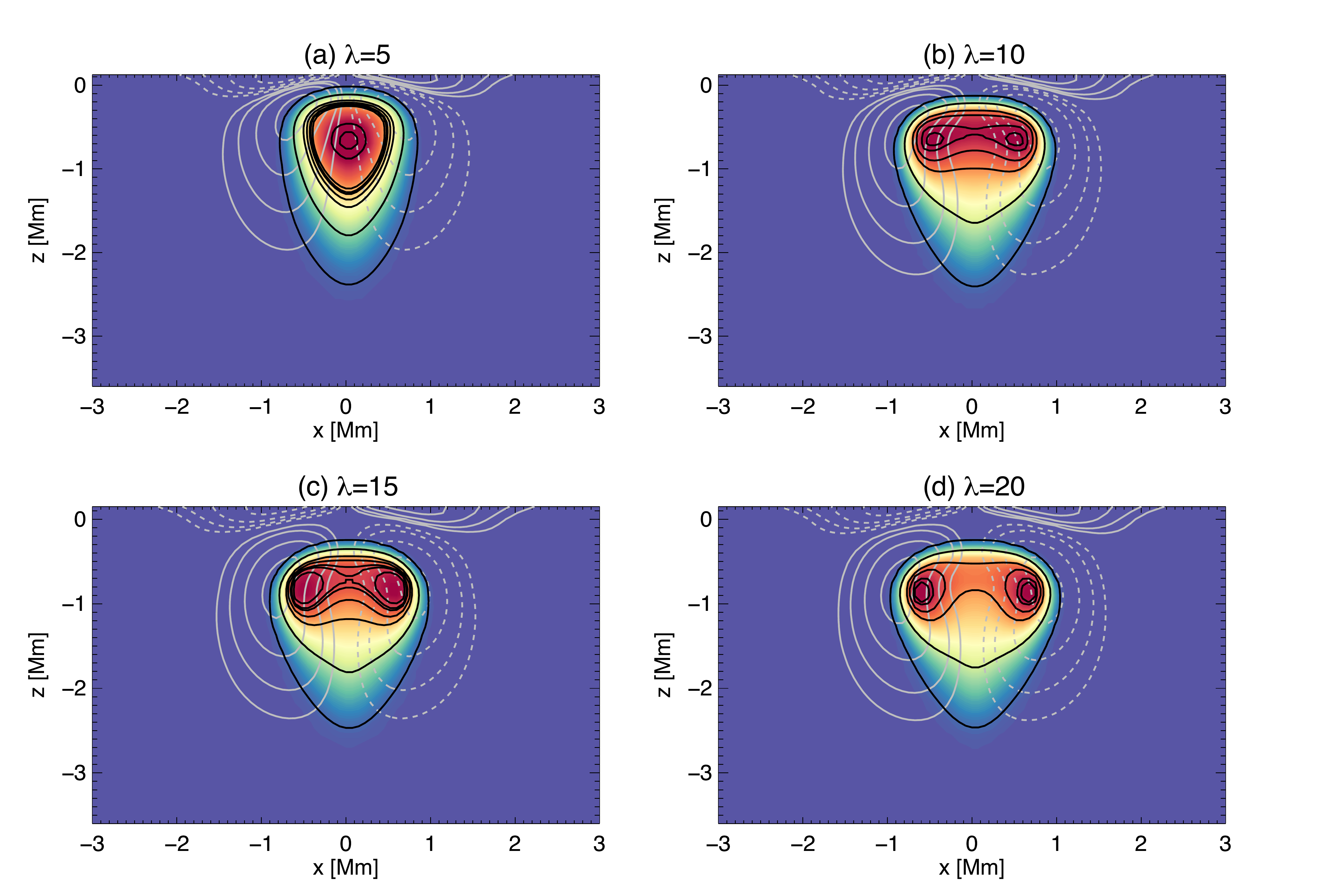}
 \caption{Cross sections of $P_{mag}$ at the $xz$-midplane for \textbf{(a):} $\lambda=5$, \textbf{(b):} $\lambda=10$, \textbf{(c):} $\lambda=15$, \textbf{(d):} $\lambda=20$. Purple is zero and dark red is $3.6\times10^4$~erg~cm$^{-3}$. Black lines are $P_{mag}$ contours (same contour values for all images). Gray contours are the $y$-component of vorticity (same contour values for all images). Solid lines are positive values and dashed lines are negative values.}
 \label{fig:cross_sections}
\end{figure*}

In the following, we firstly show the differences between the two $\lambda$ cases when both flux tubes are located in the solar interior. In 
our experiments we apply a density deficit perturbation, which is practically controlled by $\lambda$ (Eq. \ref{eq:deficit}).
In order to study how different $\lambda$ affects the initial emergence from the solar interior to the photosphere, we examine the cross section of the flux tubes at the vertical $xz$-midplane (i.e. $y=0$). There, both flux tubes are equally buoyant (same $\Delta\rho$ in Eq.~\ref{eq:deficit}, since the pressure balance terms are the same for both cases).

At $t=0$~min, the two flux tubes have equal forces applying to them (at $y=0$) along the vertical $z$-axis.
In Fig. \ref{fig:tube_center}a,b we plot for $\lambda=10$ and $\lambda=5$ respectively the $z$-component of the 
forces along the $z$-axis at $t=3$~min. We find that: (i) the gas pressure force minus the gravitational force (green line) and (ii) the 
magnetic pressure force (orange line) are approximately equal in both cases. On the other hand, the magnetic tension force (red line) 
shows a distinctive difference from the beginning of the emergence. For $\lambda=5$, the tension is higher because 
the apex of the buoyant part of the tube is more convex. Thus, the lower tension case experiences higher total force (black), 
making the flux tube to reach the photosphere faster than the one with higher tension and lower total force. Indeed, by comparing the apex positions of the flux tubes, we find that the $\lambda=10$ flux tube reaches the photosphere at $t=26$~min while the $\lambda=5$ at $t=34$~min. 
Fig. \ref{fig:tube_center}c shows the temporal evolution of the $z$-component of the magnetic tension at the center of the flux tubes, until the flux tubes reach the photosphere. We find that the tension is up to 3 times higher for smaller $\lambda$ (dashed line) throughout the initial emergence.
As a result, the high tension of smaller $\lambda$ continuously decelerates the emergence of the flux tube.

This faster emergence of the higher $\lambda$ flux tube, causes it to expand more in the vertical direction. Therefore, it leads to a faster drop of its gas and magnetic pressure in comparison to that of $\lambda=5$. In Fig. \ref{fig:tube_center}d, we plot the magnetic pressure $P_m$ at the center of the flux tubes. We plot this quantity until the moment each flux tube reaches the photosphere. 
Indeed we find that for $\lambda=5$ (dashed line), the flux tube has higher magnetic pressure at its center than for $\lambda=10$ (solid line).

Figures \ref{fig:interior}a and Fig. \ref{fig:interior}d show the magnetic pressure at the $xz$-midplane (for $\lambda=10$ and $\lambda=5$, respectively) when the flux tubes reach the photosphere (at different times). The solid white lines indicate the vertical and horizontal directions through the flux rope's center.
As we showed before, the faster vertical emergence leads to lower internal pressure for the 
higher $\lambda$ flux tube. This is apparent again by examining the $z$ and $x$ components of the forces 
along the vertical and horizontal direction through the center of the flux tube (Fig. \ref{fig:interior}b,c for $\lambda=10$ and 
Fig. \ref{fig:interior}e,f for $\lambda=5$).

Along the vertical direction, the comparison of the forces (Fig. \ref{fig:interior}b,e) shows again that the magnetic tension 
is higher for $\lambda=5$. We find that both the gradient of gas (green line) and magnetic pressure (orange line) are lower for $\lambda=10$, indicating the faster drop of pressure due to the faster emergence.

The comparison of the horizontal expansion of the flux tubes (dashed lines in Fig.~\ref{fig:interior}a,d) indicates that $\lambda=10$ expands more. Indeed, the $x$-component of the total force in the horizontal direction (black line Fig. \ref{fig:interior}c,f) shows that total force is higher for $\lambda=10$.
The forces that mainly act on the flux tube in the horizontal direction are the outward magnetic pressure, trying to expand the flux tube, and the inward gas pressure. 
In our initial flux tube, we have no twist, so initially there is no tension in the transverse direction. Still some inward tension is developed (red lines) due to the expansion of the flux tube and possibly due to internal reconnection of the field lines during this initial emergence.

An important result arises from the interior of the flux tubes during the initial emergence. Notice the contours 
in Fig. \ref{fig:interior}a,d. It is apparent that the $\lambda=5$ case emerges coherently. On the other 
hand, the $\lambda=10$ case develops two minor lobes in its interior. The lobes are formed as the center of the 
emerging flux rope starts to split in two parts. 
This also affects the plasma density in these regions. Density becomes lower in the vicinity of the 
lobes (due to higher magnetic pressure) and higher in the center of the flux rope (where the magnetic pressure drops). As a result 
we find outward directed gas pressure force (i.e. towards the lobes) in the interior of the $\lambda=10$ flux 
tube (around $-0.4$~Mm<$x$<$0.4$~Mm, Fig. \ref{fig:interior}c).

A similar behavior has been reported in previous studies of emerging flux tubes \citep[e.g.][]{Schuessler_1979,Longcope_etal1996,Moreno-Insertis_etal1996}. 
Those studies showed that the cross section of rising flux tube was distorted through the interaction with the background non-magnetized medium. More precisely, 
they found that the emerging flux tube displaced the plasma above it and forced it to flow from the apex to the flanks of the tube. This created a 
drag that deformed the cross section of the flux tube and eventually created two counter-rotating vortices. 
In our numerical simulations (e.g. $\lambda=10$ case) the appearance of the two minor lobes just before the flux rope reaches the photosphere (Fig. \ref{fig:interior}a). On the other hand, it seems that in the $\lambda=5$ case this lobe formation is inhibited and the flux rope emerges as a whole, without 
distortion (Fig. \ref{fig:interior}d).

\begin{figure*}
	\centering
	\includegraphics[width=0.9\textwidth]{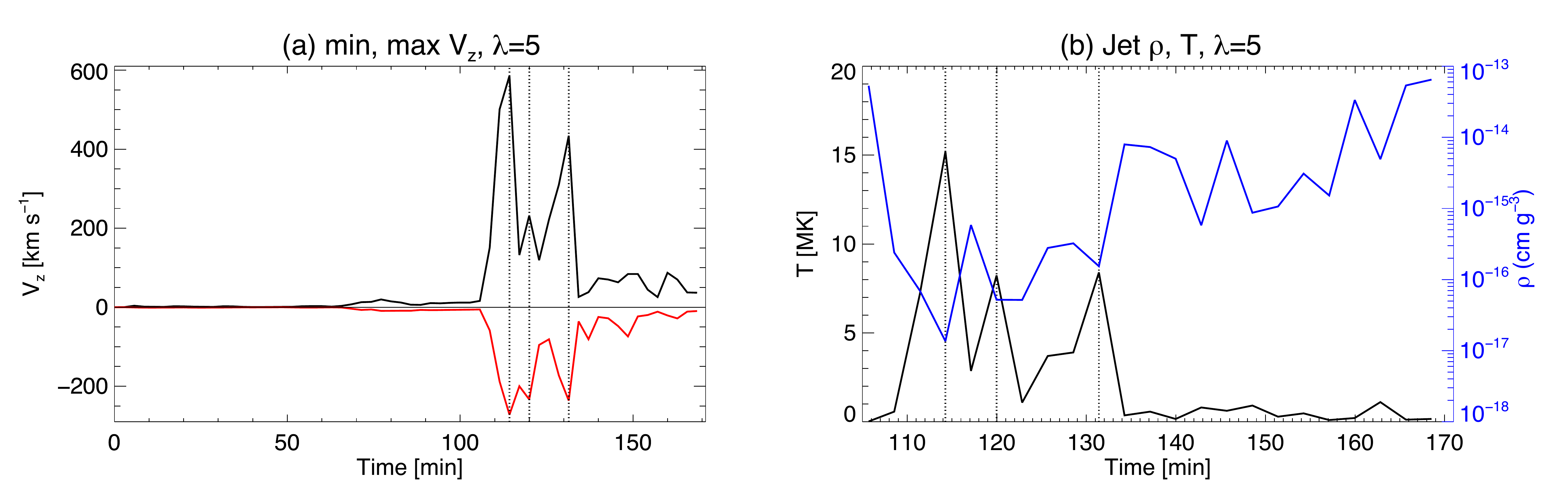}\;
	\caption{\textbf{(a)}: Temporal evolution of maximum (black) and minimum (red) $V_z$ above the photosphere. 
The vertical dashed lines correspond to the plasma ejections at $t=114,\ 120,\ 132$~min. \textbf{(b)}: Maximum 
temperature and density around the locations of maximum $V_z$.
	}
	\label{fig:vz_temp_rho_5}
\end{figure*}

To further investigate the possible role of $\lambda$ in the coherence of the flux tube during the emergence in the solar interior, we also performed simulations for $\lambda=15$ and $\lambda=20$.
In Fig. \ref{fig:cross_sections} we plot the magnetic pressure of the cross section of the flux tubes 
on the $xz$-midplane before the flux tubes reach the photosphere, for each $\lambda$. 
The black contours show $P_{mag}$. To do a closer comparison between the different $\lambda$ cases, we show the same value of $P_{mag}$. 
The contours of $P_{mag}$ indicate that lower $\lambda$ results in more coherent flux tubes that carry higher magnetic pressure flux 
at the photosphere (see dark red color inside the flux tubes). Also, we find that flux tubes  with higher $\lambda$ are more 
susceptible to splitting into two lobes. Indeed, the $\lambda=20$ case (Fig. \ref{fig:cross_sections}d) develops two very distinctive lobes, while in the  $\lambda=10$ (Fig. \ref{fig:cross_sections}b) those lobes are marginally separated (the dark red region is more elongated rather than separated).
This splitting is caused (in a similar manner to the afore-mentioned studies) by plasma flow from the apex to the flanks. This 
flow creates an aerodynamic gas pressure difference below the regions where we find the maximum horizontal expansion (around $z=-1$~Mm). 
Then, the pressure difference drives plasma motions.
To illustrate this, we overplot the contours of the $y$-component of vorticity (gray contours in Fig. \ref{fig:cross_sections}). 
We indeed find two counter rotating plasma flows along the flanks of the flux tubes. We note that the same vorticity values (for instance the second contour from inside out) are located deeper into the flux rope for 
larger $\lambda$. This signifies the presence of stronger gas-pressure driven plasma flows, which eventually deform the 
flux tubes. We find that lower tension along the flux tube axis (higher $\lambda$) results in more pronounced splitting of the flux tube. 
We should also mention (not shown here) that the flux tubes with $\lambda=15$ and $\lambda=20$ failed to emerge above the photosphere during the running time of our simulations.

We now focus on the $\lambda=5,10$ cases, which eventually emerge above the photosphere. We have found that the higher tension 
flux tube reaches the photosphere later and without lower expansion. Consequently, the flux tube with higher tension reach the photosphere with higher magnetic pressure (i.e. higher magnitude of magnetic field) and lower plasma $\beta$.
This has very important consequences on flux emergence at and above the solar surface.
We find that the $\lambda=5$ flux tube satisfies the instability criterion (Eq. \ref{eq:instability_criterion}) earlier due to its 
higher magnitude of magnetic field and the lower $\beta$. Since the emerging field is non-twisted, it spreads horizontally 
and it becomes buoyantly unstable at two regions (similar to the $\lambda$ = 10 case), forming two bipoles. 
A main difference compared to the $\lambda$ = 10 case is that these bipoles appear at the photosphere 
earlier, (at $t=66$ min, $t=154$ min for $\lambda = 10$) and at a shorter
distance (at $y=\pm1.8$~Mm, $y=4.9$~Mm for $\lambda=10$).

In the following sections we are going to highlight how the higher tension tube eventually goes through a more 
dynamic emergence and transfers more Poynting and axial flux above the photosphere than the lower tension tube.

\subsection{Jet and associated plasma dynamics}

\begin{figure*}
 \centering
 \includegraphics[width=0.8\textwidth]{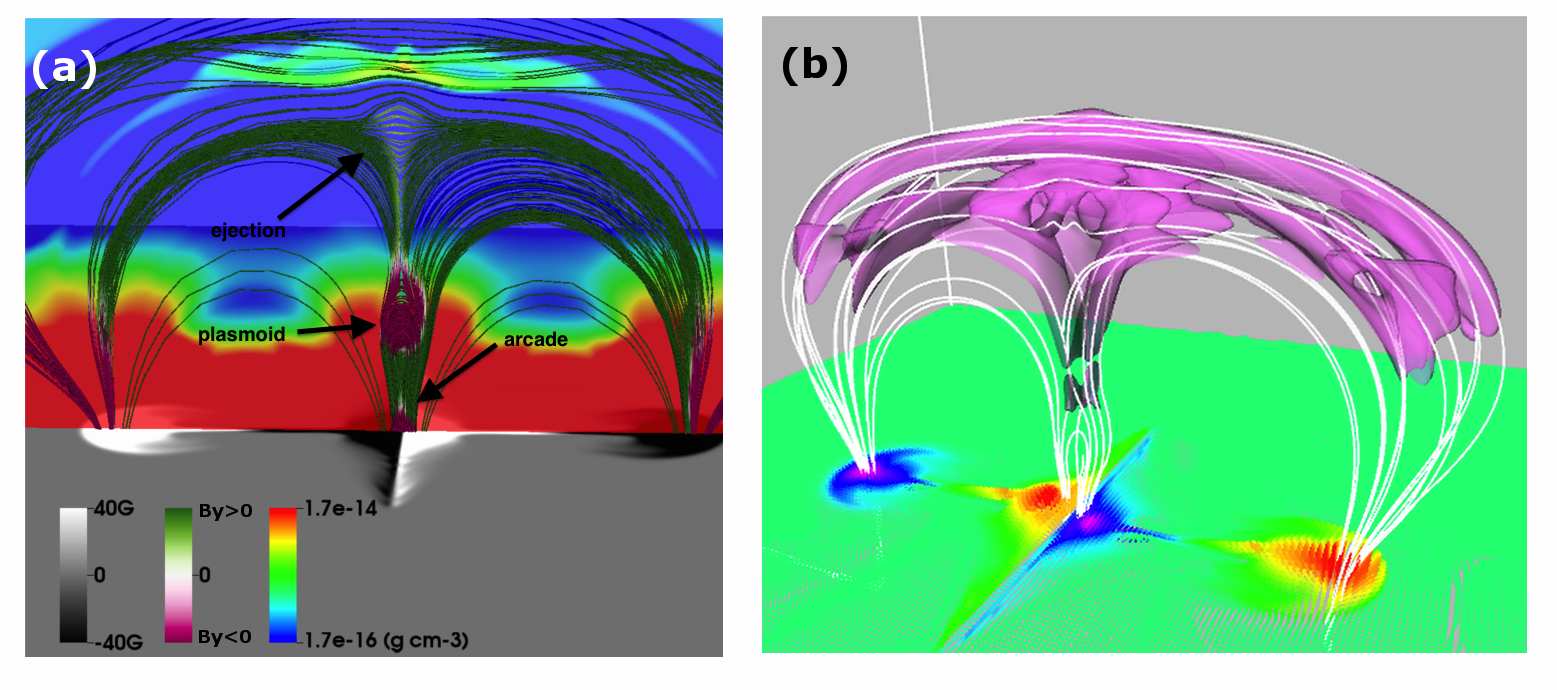}\;
 \caption{\textbf{(a)}: Fieldline topology at the vertical $yz$-midplane, showing the plasmoid(s) at the interface (at $t=108$ min). 
The horizontal cut shows the $B_z$ component of the magnetic field at the photosphere, and the purple-green colorscale shows density in the $yz$-midplane. The fieldlines are colored according to the values of $B_y$.
 \textbf{(b):} Isosurface of $V> 60\ \mathrm{km \ s^{-1}}$ showing the $t=114$~min jet. White fieldlines show the stretch of the envelope field by the ejection.  Overplotted is the full magnetic vector colored by the $B_z$ component of the field at $z=0.9$~Mm.
  }
  \label{fig:plasmoid5}
\end{figure*}

The dynamics of the jets depends on the magnetized plasma properties at the interface. We find that for $\lambda=5$, the gradient of the magnetic 
field across the interface is much higher. The resulting current sheet is 3-5 times stronger. This leads to different characteristics 
(e.g. higher plasma temperatures) for the produced plasma jets. 

Fig.~\ref{fig:vz_temp_rho_5}a shows the emission of three high-speed ($200-600$~\kms) bi-directional jets from the interface, followed by lower speed jets 
($< 100$~\kms after $t\approx 135$~min). The {\it absolute} values of the speeds of these jets are, on average, higher than in the $\lambda=10$ case. 
This is due to the higher field strength (and lower plasma $\beta$) of the magnetized material within and around the interface. However, it is 
worthwhile mentioning that in both cases the speeds of the reconnection jets are comparable to the local Alfv\'{e}n speed, as expected from the 
theory of reconnection. In this manner, we find that the speeds of the jets in the two cases are similar (determined by the Alfv\'{e}n speed).

Regarding the heating due to reconnection-driven jets, the plasma of the upward jets has a higher 
temperature in the $\lambda=5$ case. This is shown in 
Fig.~\ref{fig:vz_temp_rho_5}b, which illustrates the temporal evolution of the maximum temperature 
at the sites of maximum $V_z$.
The temperature can reach values up to $15$~MK (first jet). The following jets have progressively lower temperatures. 
In general, the temperature increases by a factor of $1/\beta$. As we discussed earlier, plasma-$\beta$ around the reconnection site is higher 
when $\lambda$ is smaller. In the time period after the onset of the jets (i.e. $t>240$~min for $\lambda=10$, $t>105$~min for $\lambda=5$), we 
calculated that (on average) $\beta\approx 0.015$ for $\lambda=5$ and $\beta\approx 0.17$ for $\lambda=5$. This 
explains the differences in the temperature of the jets when $\lambda$ is varied.

The corresponding density evolution shows a very good correlation with the temperature variations (maxima of temperature and minima of density 
occur at the same time). In comparison to the $\lambda =10$ case, the density variations lie (on average) in the same regime 
$10^{-16}-10^{-14} \ \mathrm{g \ cm^{-3}}$. Thus, in both cases, the jets bring dense plasma into the solar atmosphere.

\subsection{Fieldline topology and 3D structure of the jets.}

In both cases, the process that leads to the onset of the jets is similar. Namely, the interaction of the two bipoles leads to the formation of a strong 
current layer at the interface. The current layer becomes subject to the tearing-mode instability and plasmoid-induced reconnection leads to the emission 
of hot and high-speed jets. Hereafter, to avoid repetition, we describe briefly only the topology during the first jet. 

Figure \ref{fig:plasmoid5} shows the fieldline topology at $t=108$ min, a few minutes before the emission of the first jet. 
The vertical midplane shows plasma density (blue-red colorscale), and the horizontal slice is $B_z$ at the 
photosphere (black-white). We note that now the fieldlines are colored by the $B_y$ component of the magnetic field.
In our non-twisted configuration, the succession of positive and negative $B_y$ highlights the locations of the plasmoid(s) and the arcade. 
At this time, one plasmoid has already been ejected upwards and another one has been formed at lower heights within the interface. The ejected 
plasmoid has collided with the envelope field and, consequently, it undergoes severe deformation. The ejection of the plasmoid is confined by 
the downward tension of the envelope fieldlines. The other plasmoid is ejected soon thereafter and it is also trapped by the closed 
ambient fieldlines. 

After the ejection of the plasmoids, several jets are emitted from the interface. Fig. \ref{fig:plasmoid5}b shows the 3D structure of the first jet at 
$t=114$ min. The upward reconnection jet adopts a V-like shape (similar to the ${\lambda = 10}$ case). During its emission, the 
fast jet stretches the ambient field vertically but it does not break out. The jet is diverted into two lateral flows, which are running 
along the reconnected fieldlines of the envelope field. This leads to an apparent heating of the envelope fieldlines that join the outer polarities of the system. 

Just underneath the upward V-like jet, there is another fast jet, which is less extensive and it adopts an inverse V-like configuration. 
It extends downwards until the top of the arcade, which is formed by the O-like fieldlines that are rooted to the minor polarities of the 
system. The jet compresses the plasma at the top of the arcade and heats it to $\approx 15$~MK. The further evolution of the system 
is {\it qualitatively} similar to the $\lambda = 10$ case. For instance, the emission of several plasmoids from the interface leads to the 
onset of a series of jets, which can exist in parallel (e.g. like in Fig.~\ref{fig:mag10}f). Also, we find that the PIL undergoes deformation, developing 
horizontal undulations, around the center of the EFR, after $t=119$~min.

\subsection{Energy Transfer and Magnetic Flux}
\label{sec:energy_transfer_5}

\begin{figure*}
 \centering
 \includegraphics[width=0.9\textwidth]{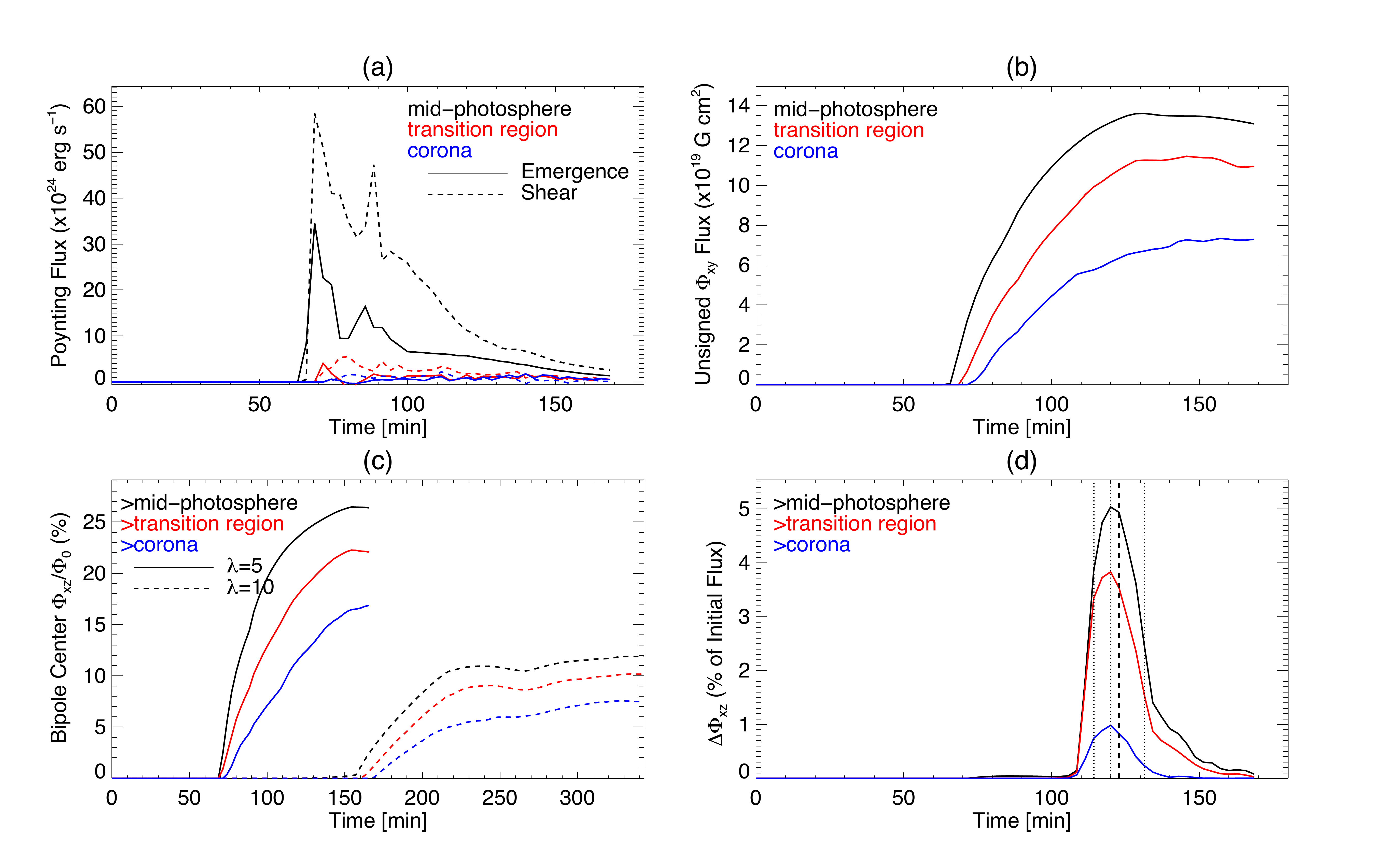}
 \caption{ \textbf{(a):} Temporal evolution of the Poynting Flux. Solid lines show the emergence term,  $F_{emergence}$, and dashed lines show the shearing term, $F_{shear}$. \textbf{(b):} The unsigned magnetic flux $\Phi_{xy}$. \textbf{(c):} The axial flux for the bipole's center in the positive $y$-direction (bipole AB) (solid). Dashed lines stand for the axial flux of $\lambda=10$. \textbf{(d):} Flux difference $\Delta\Phi_{xz}$. The dashed vertical line indicates then beginning of the undulation of the PIL 
and the dotted vertical lines indicate the times of the onset of high-speed jets from the interface.}
 \label{fig:fluxes_5}
\end{figure*}

In Fig. \ref{fig:fluxes_5}a, we plot the shearing (dashed) and emergence (solid) term of Poynting flux, above the mid-photosphere (black), transition region (red) and the corona (blue). Energy is injected above the photosphere after $t=60$~min, much sooner than the $t=140$~min of $\lambda=10$, due to the faster emergence of the flux tube above the photosphere.
One difference between the two case is that more energy is injected into the atmosphere in the smaller $\lambda$ case. Indeed, the emergence term in $\lambda=5$ is 4.2 times higher that the one in $\lambda=10$, while the shearing term is 2.4 times higher. 
This is because the emerging field with larger tension (i.e. $\lambda$=5 case) brings stronger 
magnetic field at the photosphere and evolves/expands more dynamically (e.g. faster) in the solar 
atmosphere, as we showed in the previous sections. As a result, more flux/energy is transfered into 
the outer atmospheric layer, when the subphotospheric emerging field has larger tension. Therefore, our 
experiments reveal that this specific parameter of the subphotospheric emerging field is important 
towards understanding the dynamics of the emergence at and above the solar surface.

The unsigned flux, $\Phi_{xy}$ (Fig. \ref{fig:fluxes_5}b), indicates that more flux is injected into when $\lambda$ is small. 
The maximum of the flux above the mid-photosphere (black), transition region (red) and corona (red) is 1.8 times larger the 
corresponding flux of $\lambda=10$ .
We also find evidence of flux cancellation at the photosphere (after $t=120$~min).
We find a less profound decrease of flux above the transition region, while it is not clearly apparent above the corona. In 
Fig. \ref{fig:fluxes_5}c we plot the normalized axial flux of one bipole (the bipole that is located in the positive $y$-direction) 
above different atmospheric heights (solid lines). For comparison, we plot with dashed lines the $\lambda=10$ values above the same heights. 
Above the mid-photosphere, we find that 26\%\ of the initial axial flux emerges and from it 16\%\ is located above the corona. Adding the axial flux of the two bipoles, we find that 52\%\ of the initial axial flux emerges above the photosphere, which is 1.5 times the $\lambda=10$ axial flux. Above the corona, we find 32\%\ of the axial flux to emerge, which is 2.4 times larger than in the $\lambda=10$ case.

These differences are caused by the emergence of stronger magnetic field in the lower $\lambda$ (higher tension) case. 
This results in transfer of more flux and energy into the solar atmosphere. However, we note that in both cases $2/3$ of the initial axial 
flux remain below the corona and only half of the axial flux emerges above the photosphere. Therefore, we find that in the 
cases under study a large amount of axial flux either stays at the photosphere or even submerges.

We also calculate the flux difference $\Delta\Phi_{xz}$. Above the photosphere (Fig. \ref{fig:fluxes_5}d, black line), $\Delta\Phi_{xz}$ increases rapidly after the first ejection (dotted vertical line). Its maximum, at $t=120$~min stands for 5\%\ of the initial flux being transformed into negative $B_y$. This flux difference contains both the arcade and the plasmoid's $B_y$. 
During the upward motion of the plasmoid(s), some of the plasmoid's lines reconnect with envelope fieldlines. Part 
of this flux is added to the arcade, contributing positive to $\Delta\Phi_{xz}$, and part of it enhances the envelope 
field, contributing negative to $\Delta\Phi_{xz}$. We find that overall, $\Delta\Phi_{xz}$ decreases as the plasmoid(s) moves upwards. 
The vertical dashed line indicates the start of the deformation of the PIL. 
Again, as in $\lambda=10$, $\Delta\Phi_{xz}$ decreases as the arcade follows the PIL undulation.
As we mentioned in sec. \ref{sec:energy_transfer}, $\Delta\Phi$ indicates the transformation of flux through 
reconnection. Thus, we find increased reconnection along the lobe interface in the smaller $\lambda$ case.
%
\section{Summary and Discussion}
\label{sec:conclusions}

In the present work we studied the emergence of a non-twisted flux tube in a highly stratified atmosphere.
The parameter we varied in our experiments was the buoyant part of the flux tube ($\lambda$).

We examined the initial emergence stage (below the photosphere). We found that the magnetic tension of the emerging part of the rising flux tube 
(smaller tension for higher $\lambda$) has a significant effect on the emergence process. More precisely, the $\lambda =10$ tube is distorted 
into two smaller lobes. For $\lambda =5$, the tube emerges in a more coherent manner. Moreover, smaller tension leads to higher upwards total force. 
As a result, in the $\lambda=10$ case, the tube reaches the photosphere earlier and expands more horizontally, while 
it is still within the subphotospheric region. This faster emergence/expansion  drops the flux tube's internal magnetic 
pressure faster $\lambda=10$. Subsequently, the higher $\lambda$ (smaller tension) case reaches the photosphere having smaller magnitude 
of the magnetic field. Eventually, this difference in the magnitude of the magnetic field, makes the smaller 
$\lambda$ (higher $|B|$) to emerge above the photosphere faster, although it reaches the photosphere later than  the higher $\lambda$ case.
In both cases, the initial flux tube becomes buoyant in two positions along the tube's length, forming two aligned 
bipoles at the photosphere. Since the $\lambda=5$ emerges first, the horizontal distance between the two 
bipoles at the photosphere is smaller. 

Simulations with $\lambda=15,20$ indicated that distortion becomes more intense for higher $\lambda$. 
The origin of this distortion seems to be similar to the ones described by e.g. \citet{Schuessler_1979,Longcope_etal1996,Moreno-Insertis_etal1996}. Flows at the flanks of the emerging 
flux tube increase the plasma vorticity locally. It is likely that the drag forces at the flanks of the flux tube cause the distortion. 
Although the results presented here are preliminary, we do find indications that the convex apex of the emerging flux tube 
helps the magnetic field to rise coherently. A full analysis with different magnetic field strengths, initial 
depths of the flux tubes and various twists is necessary for further conclusions.

The magnetic bipoles interact and reconnect at the interface between them, forming an envelope field and an arcade.
We find that the interface current sheet is $\approx5$ times higher when $\lambda=5$ which leads to more effective (faster) reconnection. 
From the interface current sheet, cool plasmoids are initially ejected, 
followed by several reconnection jets which are running with speeds comparable to the local Alfv\'{e}n speed.
All upward jets are confined by the envelope field and adopt a V-like shape. The downward jets compress the plasma at the top of the arcade 
and heat it to a few MK. The heating occurs in an intermittent manner due to the parallel emission of the jets at the interface. 
A similar process has been reported in more realistic experiments regarding the onset of nano-micro flares by \citet{Archontis_etal2014}.

We have also found that the value of $\lambda$ has implications on the energy and flux input into the atmosphere.
Indeed, the emergence and shearing term of the Poynting flux are 4.2 and 2.4 times higher in $\lambda=5$.  The former is due to the faster emergence and the latter is due to more pronounced horizontal flows (along the PIL) at the interface. The vertical flux transfer is $1.8$ times larger in $\lambda=5$. In $\lambda=5$ ($\lambda=10$), 52\% (30\%) of the initial axial flux is transfered above the photosphere, and 24\% (15\%) above the low corona.
The increased energy and flux transfer for smaller $\lambda$ occurs as in this case stronger magnetic field is brought at the photosphere. This causes an overall faster and more dynamical evolution in the solar atmosphere. 

In our aligned bipole configuration, we noticed that after the plasmoid ejections the PIL starts to gradually deform.
The eruptions induce a total pressure deficit (at various locations) along the PIL. The flows towards the low-pressure regimes deform the PIL, forming finger-like 
features that are mainly oriented along the $y$-axis (transverse to the PIL). This makes the post-reconnection arcade to 
become a highly curved structure consisting of twisted fieldlines.

Previous studies that compared cases with different $\lambda$ \citep[e.g.][]{Manchester_etal2004,Archontis_etal2010} showed that 
the buoyant part of a {\it twisted} flux tube affects the eruption of newly formed flux ropes, which are triggered by shearing and 
reconnection within the emerging flux region. In the present study, we investigated the effect of $\lambda$ on the emergence of 
a {\it non-twisted} field. Although the initial magnetic field configuration and initial conditions in our experiments are idealized, 
the configuration of the emerging field at the photosphere and the subsequent dynamical evolution become very complex. The emission of 
several jets from the same interface and the intermittent heating of the plasma at different atmospheric heights is just one example 
of the arisen complexity. Therefore, this parametric study showed that the variation of only one parameter of the system can affect the 
dynamics of the system considerably. More work is needed to investigate the dynamics of the jets and the atmospheric response to the 
emergence of magnetic fields from the interior of the Sun.

\begin{acknowledgements}
The authors would like to thank the anonymous Referee and the Editor for their valuable comments which helped to improve the manuscript.
The authors acknowledge support by the EU (IEF-272549 grant) and the Royal Society. The present
research has been co-financed by the European Union (European Social
Fund-ESF) and Greek national funds through the Operational Program
``Education and Lifelong Learning'' of the National Strategic Reference
Framework (NSRF) - Research Funding Program: Thales. Investing in
knowledge society through the European Social Fund.
This research has also been carried out in the frame of the research
program of the RCAAM of the Academy of Athens and has been co-financed by
the Program ``IKY Scholarships'' of the Greek national funds through the
Operational Program “Education and Lifelong Learning” of the NSRF through
the European Social Fund of ESPA 2007-2013.
Finally, the work reported in this article was additionally supported by
the SOLARNET
project, funded by the European Commision’s FP7 Capacities Program, under
the Grant Agreement 312495. The simulations were performed on the STFC and SRIF funded UKMHD cluster, at the University of St Andrews.
\end{acknowledgements}

\bibliographystyle{aa}
\bibliography{bibliography}

\end{document}